\documentclass[journal]{IEEEtran}

\usepackage{graphics} 
\usepackage{epsfig} 
\usepackage{amsmath} 
\usepackage{amssymb}  
\usepackage{cite}
\usepackage{bm}
\usepackage{subfigure}
\usepackage{acronym}
\usepackage{paralist}
\usepackage{float}
\usepackage{epstopdf}
\usepackage{color}

\newtheorem{switched stable}{Theorem}
\newtheorem{my switched}[switched stable]{Theorem}
\newtheorem{noChatteringZeno}[switched stable]{Theorem}
\newtheorem{my switched N}[switched stable]{Theorem}
\newtheorem{noChatteringZenoN}[switched stable]{Theorem}
\newtheorem{switched stable generalized}[switched stable]{Theorem}
\newtheorem{Lyapunov-like}{Definition}
\newtheorem{generalized Lyapunov-like}[Lyapunov-like]{Definition}
\newtheorem{stable1}{Lemma}
\newtheorem{barrier condition}[stable1]{Lemma}
\newtheorem{last resort}[stable1]{Lemma}
\newtheorem{Corollary}{Corollary}
\newtheorem{Remark}{Remark}
\newtheorem{Problem}{Problem}
\newtheorem{Assumption}{Assumption}

\DeclareMathOperator{\atan2}{atan2}

\DeclareMathOperator{\F}{\mathrm F}

\DeclareMathOperator{\R}{\mathbb R}

\DeclareMathOperator{\N}{\textrm N}

\begin{document}

\title{Decentralized Goal Assignment and Safe Trajectory Generation in Multi-Robot Networks via Multiple Lyapunov Functions}

\author{Dimitra Panagou, Matthew Turpin and Vijay Kumar 
\thanks{Dimitra Panagou is with the Department of Aerospace Engineering, University of Michigan, Ann Arbor, MI, USA.  \texttt{dpanagou@umich.edu}} 
\thanks{Matthew Turpin was with the GRASP Lab, School of Engineering and Applied Science, University of Pennsylvania, Philadelphia, PA, USA when this work was done. He is now with Qualcomm. \texttt{mturpin@seas.upenn.edu}}
\thanks{Vijay Kumar is with the GRASP Lab, School of Engineering and Applied Science, University of Pennsylvania, Philadelphia, PA, USA. \texttt{kumar@seas.upenn.edu}}
\thanks{We gratefully acknowledge the support of ARL grant W911NF-08-2-0004, ONR grant N00014-09-1-1051, and TerraSwarm, one of six centers of STARnet, a Semiconductor Research Corporation program sponsored by MARCO and DARPA.}
}

\maketitle
\thispagestyle{empty}
\pagestyle{empty}

\acrodef{wrt}[w.r.t.]{with respect to}
\acrodef{apf}[APF]{Artificial Potential Fields}
\acrodef{OGA}[\textsc{oga}]{Optimal Goal Assignment}
\acrodef{CA}[\textsc{ca}]{Collision Avoidance}
\acrodef{gas}[\textsc{GAS}]{Globally Asymptotically Stable}
\acrodef{ges}[\textsc{GES}]{Globally Exponentially Stable}
\acrodef{rbf}[\textsc{RBF}]{Recentered Barrier Function}

\begin{abstract}
This paper considers the problem of decentralized goal assignment and trajectory generation for multi-robot networks when only local communication is available, and proposes an approach based on methods related to switched systems and set invariance. A family of Lyapunov-like functions is employed to encode the (local) decision making among candidate goal assignments, under which a group of connected agents chooses the assignment that results in the shortest total distance to the goals. An additional family of Lyapunov-like barrier functions is activated in the case when the optimal assignment may lead to colliding trajectories, maintaining thus system safety while preserving the convergence guarantees. The proposed switching strategies give rise to feedback control policies that are computationally efficient and scalable with the number of agents, and therefore suitable for applications including first-response deployment of robotic networks under limited information sharing. The efficacy of the proposed method is demonstrated via simulation results and experiments with six ground robots.
\end{abstract}

\section{Introduction}
Task (target) assignment problems in multi-agent systems have received great interest within the robotics and controls communities in the past couple of years, in part because they encode the accomplishment of various objectives in applications such as surveillance, exploration and coverage \cite{Yan-IJARS13}. A common thread towards solving assignment problems is the development of algorithms that assign targets to agents by optimizing a predefined criterion \cite{Kloder_TRO06, Bullo_TAC09, Liu-RSS-11,Jouandeau, Zhao-2016}. Standard goal assignment strategies are the greedy assignment \cite{Yamauchi-RAS99}, the iterative assignment \cite{Werger-Agents00}, the multiple Traveling Salesman assignment \cite{Faigl-IROS12}, the MinPoS assignment \cite{Bautin-2012} and market-based mechanisms \cite{Lagoudakis-IROS04, Choi_How_TRO09, Hawley-DARS2013}. 

The main focus in this line of research is mostly related to the optimization aspects of the problem and the associated computational complexity, so that suboptimal solutions with certain performance guarantees are derived. A hybrid control solution to multi-robot goal assignment is given in \cite{Zavlanos_TRO08}, while a game-theoretic approach is presented in \cite{Shamma_ASME07}. More recently, the problem of assigning a group of mobile robots to an equal number of static targets under the assumption that the robots have limited range-based communication and target-sensing capabilities was studied in \cite{Yu-2015}, and in \cite{Bai-2017}, respectively.  

Nevertheless, such optimization-based techniques are often derived under relaxed assumptions, such as point or static agents, which are in principle not acceptable in realistic settings. For instance, first-response or search-and-rescue missions using robotic agents (e.g., aerial robots) typically involve multiple tasks that, on one hand, may need to be performed as quickly as possible, yet on the other hand they can be accomplished only when information sharing is available, and under tight safety guarantees. In response to this challenge, in our earlier work \cite{Turpin_WAFR12, Turpin_IJRR14} we proposed a centralized approach on the Concurrent Assignment and Planning of Trajectories (CAPT) for multi-robot systems, that produces minimum distance traveled trajectories for non-point robots, that are additionally collision-free under mild assumptions on the initial locations of the robots and the locations of the goals. Hence we simultaneously addressed two challenges: \begin{inparaenum}\item[(i)] the combinatorially complex problem of finding a suitable assignment of robots to goal locations, and \item[(ii)] the generation of collision-free trajectories for every robot under given assumptions\end{inparaenum}. More specifically, we considered unlabeled (interchangeable) robots and goal locations, and proposed algorithmic solutions for the centralized CAPT problem, as well as a decentralized implementation of the centralized algorithm that produces suboptimal solutions in terms of total distance traveled. It should be noted that although the combinations of $N$ robots to $N$ goals result in intractable enumeration for all but the smallest problems, fortunately, the $\mathcal O({N}^3)$ Hungarian Algorithm from the Operations Research community \cite{Munkres57} optimally solves this problem in a centralized manner, while extensions to distributed systems have been also developed \cite{Giordani10}. An extensive treatment of assignment problems and algorithms can be found in \cite{Bertsekas-Network}. Our simulated teams of $N$ robots under the CAPT algorithm on a single computer take approximately $10^{-7}{N}^3$ seconds to compute the optimal assignment, i.e., a team of 100 robots takes about 0.1 seconds to plan its trajectories. This is efficient for the considered multi-robot applications. \emph{However, the decentralized implementation of the CAPT algorithm does not in general ensure safety for the multi-robot system, i.e., that the trajectories of the robots are collision-free}. 

Let us note that some technical ingredients of the problem considered here might be applicable or useful to other types of decentralized/distributed multi-robot planning and control problems as well, such as, for instance, robotic team reconfiguration \cite{Fitch-2016}, control of modular robots, or formation/flocking control. However, the main novel characteristics that differentiate our work from similar problems and approaches are: (i) the \emph{casting of both the assignment and safety problems for a dynamic multi-agent/multi-robot system into a switched systems stability problem}, and (ii) the development of \emph{decentralized} feedback control algorithms that generate convergent and collision-free trajectories with provable guarantees. The contributions of this work are described in more detail in the following section.

\subsection{Overview and Contributions}

In this paper we build upon our earlier work \cite{Turpin_WAFR12, Turpin_IJRR14} that considers a computationally efficient way of safely assigning goals in a multi-robot system \emph{in a centralized manner}, and we focus on the problems of concurrently assigning goals and generating collision-free trajectories for the \emph{decentralized} case, i.e., for when the robots have limited sensing and communication capabilities and hence exchange information only locally, i.e., when they become connected. We develop algorithms that concurrently address the problems of:
\begin{itemize}
\item[($P_1$)] Assigning goal locations to agents by minimizing a cost function expressing distance to the goals.
\item[($P_2$)] Designing \emph{feedback} control policies guaranteeing:
\begin{itemize}
\item[(i)] the convergence of the agents to their assigned goals,
\item[(ii)] that the resulting trajectories are collision-free.
\end{itemize}
\end{itemize}
The key specifications in the proposed formulation are:
\begin{itemize}
\item[($S_1$)] Agents and goal locations are interchangeable, which means that the mission is considered accomplished when each agent has converged to some goal location.
\item[($S_2$)] Information exchange between a pair of agents is feasible and reliable only when they lie within a certain communication range, which means that the decision making on the optimal goal assignment (in terms of distance to the goals) can be performed only locally, i.e., in a decentralized fashion.
\item[($S_3$)] Agents are modeled as non-point robots.
\end{itemize}

The adopted decentralized formulation and the underlying goal assignment and control design techniques in the current paper offer several improvements compared to the earlier work of the authors' in the conference paper \cite{Panagou_Turpin_Kumar_ICRA14}, namely: The conservative conditions on the initial positions of the agents and of the goal locations in \cite{Panagou_Turpin_Kumar_ICRA14} are now relaxed via the design of vector-field controllers instead of gradient-based controllers. The proposed vector fields are always well-defined everywhere in each agent's state space, opposed to the barrier functions utilized in \cite{Panagou_Turpin_Kumar_ICRA14} that exhibit some pathological situations when the distances between neighbor agents and goal locations become smaller than a certain distance. This way, the overall switched system can be implemented by forcing the agents switch among \emph{two} control policies (described below), instead of \emph{three} in \cite{Panagou_Turpin_Kumar_ICRA14}, where the third controller was taking care of the pathological situations of the barrier functions. In addition, the initial positions of the robots can be selected to be $R$ apart, compared to the conference version that required them to be $2R\sqrt{2}$ apart. More specifically, the novelties of the proposed approach are as follows: 

\textit{Contribution 1:} We consider the goal assignment and trajectory generation problem posed in \cite{Turpin_IJRR14} that seeks to minimize the distance traveled by all robots, and formulate it into a switched systems framework \cite{Liberzon}; this provides a novel and natural way of encoding multiple tasks via Lyapunov-like functions, and obtaining guarantees on the accomplishment of the encoded control objectives. We build our approach based on multiple Lyapunov-like functions \cite{Branicky_TAC1998} and set invariance \cite{Blanchini_Miani} techniques. We first encode the decision making on the optimal goal assignment (which in the sequel we call the \ac{OGA} policy) as a state-dependent switching logic among a family of candidate Lyapunov-like functions. Each Lyapunov-like function encodes a cost under a candidate goal assignment, that is, the sum of squared distances to the goals. The switching logic dictates that, when a set of agents becomes connected at some time instant $t$, they decide to switch to the Lyapunov-like function of minimum value at time $t$. We show that this decision making gives rise to a \ac{gas} switched system which furthermore does not suffer from Zeno trajectories.
Then, based on our recent work in \cite{Panagou_TAC14, Panagou_submitted}, we build an additional state-dependent switching logic which employs a family of Lyapunov-like barrier functions and vector fields encoding both inter-agent collision avoidance and convergence to the goal locations determined by the \ac{OGA} policy. This control policy (in the sequel called the \ac{CA} policy) provides sufficient conditions on determining whether the \ac{OGA} policy is safe, and furthermore serves as a supervisor that takes action only when safety under the \ac{OGA} policy is at stake. We show that the switching between the \ac{OGA} policy and the \ac{CA} policy results in safe and almost globally attractive trajectories for the multi-robot system, at the expense of possibly resorting to suboptimal paths in terms of total distance traveled; this situation appears only in the cases when the \ac{CA} policy forces the agents to deviate from their optimal (straight line) paths to the goals, in order to maintain system safety. To the best of our knowledge, the proposed approach of encoding assignment and motion tasks via Lyapunov-like functions is a novel approach in multi-robot task and motion planning, and introduces a framework of correct-by-construction switched (hybrid) systems that satisfy the encoded tasks, along with the \emph{feedback} control policies that accomplish the encoded tasks. 

\textit{Contribution 2:} Part of this work appeared in \cite{Panagou_Turpin_Kumar_ICRA14}. Compared to the conference paper, here we develop a different conflict resolution and collision avoidance strategy that is based on directly defining feedback vector fields for each one of the agents, instead of barrier functions. The method generates safe reference directions and feedback controllers for the agents, and alleviates \begin{inparaenum}\item[(i)] the issues often raised in pure gradient-based solutions, where the behavior of the gradient-based controller is hard to tune, as well as \item[(ii)] the use of the third family of Lyapunov-like functions in the switching strategy in \cite{Panagou_Turpin_Kumar_ICRA14}\end{inparaenum}. Furthermore, (iii) the new design provides safety and convergence guarantees for the decentralized multi-robot system under less restrictive assumptions on the spatial distributions of the agents and the goal locations, compared to the assumptions adopted in \cite{Turpin_IJRR14, Panagou_Turpin_Kumar_ICRA14}. Finally, (iv) we provide a robustness result on the safety of the resulting multi-robot trajectories in the case of communication failures, and (v) we implement and verify the effectiveness of the proposed algorithms experimentally with six ground robots.

The paper is organized as follows: Section \ref{Problem Formulation} presents the mathematical formulation of the problem, while the adopted notions and tools from switched systems theory are reviewed in Section \ref{Switched Systems Theory}. The proposed goal assignment and collision avoidance policies are given in Sections \ref{OGA} and \ref{CA}, respectively, along with the mathematical proofs that verify their correctness. Simulation results to evaluate their efficacy are included in Section \ref{Simulation Results}, along with a comparison with the results in the conference version \cite{Panagou_Turpin_Kumar_ICRA14}, and a discussion on the communication requirements and planning times in the connected groups of agents. Experimental results with six ground robots are provided in Section \ref{Experimental Results}. Section \ref{Conclusions} summarizes our conclusions and thoughts on future research.

\section{Problem Formulation}\label{Problem Formulation}

We consider $\N$ agents $i$ and equal number of goal locations $G_i$, $i\in\mathcal N=\{1,2,\dots,\N\}$. The objective of each agent $i$ is to safely converge to some goal $G_i$, while collaborating with its locally connected agents (see definition below) so that their total traveled distance is minimized. The motion of each agent $i$ is governed by single integrator dynamics:
\begin{align}
\label{system}
\dot{\bm r}_i=\bm u_i,
\end{align}
where $\bm r_i=\begin{bmatrix}x_i&y_i\end{bmatrix}^T$ is the position vector of agent $i$ \ac{wrt} a global Cartesian coordinate frame $\mathcal G$, and $\bm u_i$ is its control vector comprising the velocities $u_{xi}$, $u_{yi}$ \ac{wrt} the frame $\mathcal G$. We denote the speed of agent $i$, i.e., the magnitude of the velocity vector $\bm u_i$, as $u_i=\|\bm u_i\|=\sqrt{u_{xi}^2+u_{yi}^2}$.

We assume that each agent $i$:
\begin{inparaenum}
\item[(i)] has access to its position $\bm r_i$ and velocities $\bm u_i$ via onboard sensors,
\item[(ii)] can reliably exchange information with any agent $j\neq i$ which lies within its communication region $\mathcal C_i : \{\bm r_i\in\R^2, \bm r_j\in\R^2 \; | \; \|\bm r_i - \bm r_j\|\leq R_c\}$, where $R_c$ is the communication range. In other words, a pair of agents $(i,j)$ is connected as long as the distance $d_{ij}=\|\bm r_i - \bm r_j\|\leq R_c$.
\end{inparaenum}
\begin{Remark}
Note that maintaining connectivity for the entire network is not fundamental for the proposed assignment and collision avoidance strategies; in other words, maintaining a connected graph is neither necessary, nor a control objective in the proposed problem and approach.
\end{Remark}
\begin{Remark}
The consideration of equal number of robots and goal locations is mainly due to the fact that we do not consider the problem of \emph{globally} assigning goals to robots at the initial time of the deployment, but rather we consider the problem of \emph{locally} re-assigning goals to robots whenever/if the robots get connected along with safety and convergence guarantees. The initial goal assignment is random and hence, suboptimal in terms of total traveled distance in general. In the case of $\textrm M$ goal locations and $\N$ robots with $\textrm M>\N$, our approach would be initialized by randomly assigning a goal location per robot, and from there on the selected $\N$ goals would remain the same over the evolution of the system (i.e., the remaining $\textrm M-\N$ would never be selected by any robot). Similarly, in the case of $\textrm M<\N$, only $\textrm M$ robots would be randomly assigned a goal location each, while the remaining $\N-\textrm M$ robots would not be assigned any goal.   
\end{Remark}

\vspace{2mm}

The task is considered completed as long as each agent converges to some goal location, with goal locations being interchangeable -- each agent can converge to any goal location. This specification defines $\N!$ possible goal assignments $k\in\{1,2,\dots,\N!\}$. Each goal assignment can be encoded via an assignment matrix: $$A_k=\{\alpha_{im}\}=\begin{bmatrix}\alpha_{11}&\ldots&\alpha_{1\N}\\\vdots&\vdots&\vdots\\\alpha_{\N1}&\ldots&\alpha_{\N\N}\end{bmatrix},$$ with the rows representing agents and the columns representing goal locations. For agent $i\in\{1,\dots,\N\}$ assigned to goal $G_m$, $m\in\{1,\dots,\N\}$, one has $\alpha_{im}=1$ and the remaining elements of the $i$-th row equal to zero. For instance, for $2$ agents and $2$ goal locations, the possible $\N!=2$ goal assignments are encoded via the assignment matrices:
\begin{align*}
A_1=\begin{bmatrix}1&0\\0&1\end{bmatrix}, \;\; A_2=\begin{bmatrix}0&1\\1&0\end{bmatrix}.
\end{align*}
In this case, the matrix $A_1$ encodes that agent $1$ goes to $G_1$ and agent $2$ goes to $G_2$, while the matrix $A_2$ encodes that agent $1$ goes to $G_2$ and agent $2$ goes to $G_1$.

Denote $\bm r_{G_m}=\begin{bmatrix}x_{G_m}&y_{G_m}\end{bmatrix}^T$ the position vectors of the goal locations $G_m$ \ac{wrt} the global frame. Then, the assigned goal position for agent $i$ under the $A_k$ assignment matrix (or under the $k$-th goal assignment) can be expressed as: $$\bm r_{g_i}^{(k)}=\sum_{m=1}^{\N}\alpha_{im}^{(k)}\;\bm r_{G_m},$$ where $\alpha_{im}^{(k)}$ is the $i$-th row of $A_k$.

The next lemma establishes that under any assignment, the robots will converge to the respective goal locations. This is essential for the switching control strategy developed later.

\begin{stable1}\label{stable}
The position trajectories $\bm r_{i}(t)$ of agent $i$ are \ac{ges} \ac{wrt} the $k$-th goal assignment under the control law:
\begin{align}
\label{controller}
\bm u_i^{(k)}=-\lambda_i\left(\bm r_i-\bm r_{g_i}^{(k)}\right),
\end{align}
where $i\in\{1,\dots,\N\}$, $\lambda_i>0$.

\begin{proof}
Consider the candidate Lyapunov function $V_i^{(k)}$ for each agent $i$ under the $k$-th assignment as:
\begin{align*}
V_i^{(k)}=\left\|\bm r_i - \bm r_{g_i}^{(k)}\right\|^2,
\end{align*}
which encodes the distance of agent $i$ from its goal $\bm r_{g_i}^{(k)}$ under the goal assignment $k$. The function $V_i^{(k)}$ is continuously differentiable, positive definite, and radially unbounded. Furthermore, the closed-loop system vector field \eqref{system} under the control law \eqref{controller} is globally Lipschitz continuous. Given that $\dot{\bm r}_{G_i}=\bm 0$, one has that the time derivative of $V_i^{(k)}$ along the system trajectories under the control law \eqref{controller} reads:
\begin{align*}
\dot V_i^{(k)}&=-2\lambda_i \left\|\bm r_i-\bm r_{g_i}^{(k)}\right\|^2\leq-2\lambda_i V_i^{(k)},
\end{align*}
yielding that the position trajectories $\bm r_i(t)$ of each agent $i$ are \ac{ges} \ac{wrt} the assigned goal $\bm r_{g_i}^{(k)}$.
Furthermore, by considering the Lyapunov function:
\begin{align}
\label{sumLyap}
V^{(k)}=\sum_{i=1}^{\N}\; V_i^{(k)},
\end{align}
encoding the sum of distances of the agents to their assigned goals, one has that \eqref{controller} render its time derivative:
\begin{align*}
\dot V^{(k)}=\sum_{i=1}^{\N}\;-2\lambda_i V_i^{(k)}&\leq -2\min_i \{\lambda_i\} \sum_{i=1}^{\N} V_i^{(k)}=\\
&=-2\min_i\{\lambda_i\} V^{(k)},
\end{align*}
which implies that the sum of squared distances of agents $i$ to their assigned goals $\bm r_{g_i}^{(k)}$ is \ac{ges} to zero.
\end{proof}
\end{stable1}

\vspace{2mm}

\noindent \emph{Problem Description:} Our goal is to develop decentralized algorithms that force $\N_c\leq \N$ connected agents to decide whether they keep their current goal assignment or swap goals. The criterion on the decision making is to choose the assignment $k\in\{1,\dots,\N_c!\}$ which \emph{minimizes the sum of total distance (i.e., results in the shortest paths) to the goal locations}. In the sequel, we refer to this assignment as the optimal assignment. The challenge in the case of non-point robots is to ensure that the optimal assignment results not only in stable, but also in collision-free trajectories \ac{wrt} the assigned goal locations. We develop a switching policy between optimal (shortest) and safe (collision-free) trajectories with provable guarantees. This policy is realized via switching between the \ac{OGA} and \ac{CA} policies defined in the sequel, while the switched systems and control framework is employed to facilitate the stability and safety analysis for the resulting trajectories of the multi-robot system. We first review some basic results on switched systems theory.

\section{Notions from Switched Systems Theory}\label{Switched Systems Theory}
Following \cite{Branicky_TAC1998}, let us consider the prototypical example of a switched system:
\begin{align}
\label{prototype}
\dot x(t)=f_k(x(t)), \quad k\in\mathcal K=\{1,2,\dots,K\},
\end{align}
where $x(t)\in\mathbb R^n$, each $f_k$ is globally Lipschitz continuous and the $k$'s are picked in such a way that there are finite switches in finite time. Consider a strictly increasing sequence of times: $$T = \{t_0, t_1, \ldots , t_n, \ldots , \}, \quad n\in\mathbb{N},$$ the set: $I(T)=\bigcup_{n\in\mathbb N}[t_{n},t_{n+1}]$ denoting the interval completion of the sequence $T$, and the switching sequence:
\begin{align*}
\Sigma = \{x_0; (k_0,t_0),(k_1,t_1),\ldots,(k_n,t_n),\ldots\},
\end{align*}
where $t_0$ is the initial time, $x_0$ is the initial state, $\mathbb N$ is the set of nonnegative integers, and the elements in the set $\{k_0,k_1,\ldots,k_n,\ldots\}$ take values in $\mathcal K$. The switching sequence $\Sigma$ along with \eqref{prototype} completely describes the trajectory of the switched system according to the following rule: $(k_n,t_n)$ means that the system evolves according to:
\begin{align*}
\dot{x}(t)=f_{k_n}(x(t),t)
\end{align*}
for $t_n \leq t < t_{n+1}$. Equivalently, for $t\in[t_{n},t_{n+1})$ one has that the $k_{n}$-th subsystem is active. We assume that the switching sequence is minimal in the sense that $k_n\neq k_{n+1}$. 

Now, for any subsystem $k \in \mathcal K$, denote:
\begin{align*}
\Sigma \; | \; k = \{ t_{0|k} , t_{1|k} , \ldots , t_{n|k} , t_{{n+1|k}} , \ldots, \}
\end{align*}
the sequence of switching times when the $k$-th subsystem is \emph{switched on} or \emph{switched off}. Hence, $$I(\Sigma\;|\;k)=\bigcup_{n\in\mathbb N}[t_{n|k},t_{n+1|k}]$$ is the set of times that the $k$-th subsystem is active. Denote $E(T) = \{t_0, t_2, t_4, \ldots , \}$ the even sequence of $T$; then, $$E(\Sigma\;|\;k) = \{t_{0|k}, t_{2|k} , \dots , t_{n|k}, \dots \;|\; n \in \mathbb{N}\}$$ denotes the sequence of the \emph{switched on} times of the $k$-th subsystem. In the sequel, the notation $t_{n|k}$ will be equivalent to $k(t_n)$, denoting the time instance $t_n$ when the $k$-th subsystem becomes active.

\begin{Lyapunov-like}\cite{Branicky_TAC1998}\label{Branicky's Theorem}
A $C^1$ function $V:\R^n\to \R^+$ with $V(0)=0$ is called a Lyapunov-like function for a vector field $f(\cdot)$ and the associated trajectory $x(\cdot)$ over a strictly increasing sequence of times $T$ if:
\begin{enumerate}
\item[(i)] $\dot V(x(t)) \leq 0$, $\forall t\in I(T)$,
\item[(ii)] $V$ is monotonically non-increasing on $E(T)$.
\end{enumerate}
\end{Lyapunov-like}
\begin{switched stable}\cite{Branicky_TAC1998}
If for each $k\in\mathcal K$, $V_k$ is a Lyapunov-like function for the $k$-th subsystem vector field $f_k(\cdot)$ and the associated trajectory over $\Sigma\; |\; k$, then the origin of the system is stable in the sense of Lyapunov.
\end{switched stable}
If in addition, for any $t_p, t_q \in E(\Sigma \; | \; k)$, where $p < q$,
\begin{align}
\label{GAS condition}
V_{k(t_q)}(x(t_{q})) - V_{k(t_p)}(x(t_{p})) \leq -W\left(x(t_{p}) \right),
\end{align}
holds for some positive definite continuous function $W$, then the origin of the system is \ac{gas} \cite{Peleties_91}.

\begin{figure}
\centering
\includegraphics[width=0.75\columnwidth,clip]{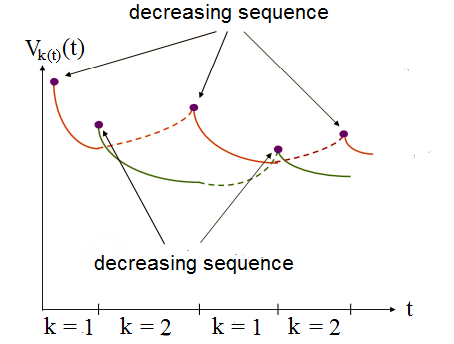}
\caption{Branicky's ``decreasing sequence" condition. Image from \cite{Zhao_slides}.}
\label{fig:decreasing sequence}
\end{figure}

The theorem essentially states that: Given a family of functions $\{V_k\; |\; k\in \mathcal K\}$ such that the value of $V_k$ decreases on the time interval when the $k$-th subsystem is active, if for every $k$ the value of the function $V_k$ at the beginning of such interval exceeds the value at the beginning of the next interval when the $k$-th system becomes active (see Fig. \ref{fig:decreasing sequence}), then the switched system is \ac{gas}.

A fundamental assumption of the theorem is the decreasing condition of $V$ on $E(\Sigma \; | \; k)$. This is quite conservative and, in general, hard to check \cite{Zhao_Automatica_2008}. The concept of generalized Lyapunov-like functions relaxes this condition.
\begin{generalized Lyapunov-like}\cite{Zhao_Automatica_2008}
A $C^0$ function $V:\R^n\to \R^+$ with $V(0)=0$ is called a generalized Lyapunov-like function for a vector field $f(\cdot)$ and the associated trajectory $x(\cdot)$ over a strictly increasing sequence of times $T = \{t_0, t_1, \ldots , t_n, \ldots , \}, \quad n\in\mathbb{N}$, if there exists a function $h : \R^+\to \R^+$ satisfying $h(0)=0$, such that: $V(x(t)) \leq h(V(x(t_{n})))$, for all $t \in [t_{n}, t_{n+1})$ and all $n\in \mathbb N$.
\end{generalized Lyapunov-like}

\begin{switched stable generalized}\label{zhao}
\cite{Zhao_Automatica_2008} Consider the prototypical switched system \eqref{prototype} and assume that for each $k\in\mathcal K$ there exists a positive definite generalized Lyapunov-like function $V_k(x)$ \ac{wrt} $f_k(x(t))$ and the associated trajectory $x(t)$, $t\in[t_n,t_{n+1})$. Then:
\begin{enumerate}
\item [(i)] The origin of the switched system is stable if and only if there exist class $\mathcal{GK}$ functions $\alpha_k(\cdot)$ satisfying
\begin{align}
\label{zhao condition}
V_k(x(t_{2q|k})) - V_k(x(t_{0|k})) &\leq \alpha_k(\| x_0 \|), \\
q \geq 1, \;\;k &= 1,2, . . . , K. \nonumber
\end{align}
\item [(ii)] The origin of the switched system is asymptotically stable if and only if \eqref{zhao condition} holds and there exists $k$ such that $V_k(x(t_{2q|k})) \rightarrow 0$ as $q \rightarrow \infty$.
\end{enumerate}
\end{switched stable generalized}

The theorem essentially states that stability of the switched system is ensured as long as $V_k(x(t_{2q|k})) - V_k(x(t_{0|k}))$, which is the change of $V_k$ between any ``switched on" time $t_{2q|k}$ and the first active time $t_{0|k}$, is bounded by a class $\mathcal{GK}$ function, regardless of where $V_k(x(t_{0|k}))$ is.

\begin{Remark}
The condition \eqref{zhao condition} is equivalently rewritten as:
\begin{align*}
\sum_{q=1}^\infty\left(V_k(x(t_{2q|k})) - V_k(x(t_{2(q-1)|k}))\right)\leq \alpha_k (\|x_0\|),
\end{align*}
for any $k \in\mathcal K$. Note that here $V_k(x(t_{{2q}|k})) - V_k(x(t_{2(q-1)|k}))$, $q\geq 1$, stands for the change of $V_k(x)$ at two adjacent ``switched on" times. Thus, the condition means that $V_k$ is allowed to grow on $E(\Sigma\; |\; k )$ but the total such change on any finite time interval should be bounded from above by a class $\mathcal{GK}$ function.
\end{Remark}
\begin{Remark}
Theorem \ref{zhao} relaxes the decreasing requirement of $V_k(x(t))$ on the corresponding active intervals. Instead, we only need that $V_k(x(t))$ on an active interval does not exceed the value of some function of $V_k$ at the ``switched on" time.
\end{Remark}

\section{The Optimal Goal Assignment Policy}\label{OGA}

\subsection{The $\N=2$ Robot Case}
For simplicity, we first consider $\N=2$ agents $i$, $j$ that need to safely move towards goal locations $G_1$, $G_2$, and build a switching logic that locally decides on the \ac{OGA}.

Assume that the agents initiate at $t_0=0$ such that $d_{ij}(t_0)>R_c$ under a \emph{random} assignment $k\in\{1,2\}$ (i.e., not necessarily the optimal one) and move under the control laws \eqref{controller}. If the inter-agent distance happens to remain always greater than the communication range $R_c$, i.e., if $d_{ij}(t)>R_c$, $\forall t\in[0,\infty)$, we say that the agents are never involved in a meeting event, and as thus, no decision on the goal assignment needs to be made. Out of Lemma \ref{stable}, the position trajectories $\bm r_i(t)$, $\bm r_j(t)$ are \ac{ges} \ac{wrt} the assigned goal locations. Clearly, in such cases, no inter-agent collisions occur.

Assume now that the agents initiate at $t_0=0$ such that $d_{ij}(t_0)\leq R_c$, or that at some time instant $t_d>0$ they lie within distance $d_{ij}(t_d)\leq R_c$ under some goal assignment $k\in\{1,2\}$ (again, not necessarily the optimal one). We say that the agents are involved in a meeting event and a decision regarding on the goal assignment has to be made. The proposed decision making strategy is defined as follows: The agents:
\begin{enumerate}
\item Exchange information on their current positions $\bm r_i(t_d)$, $\bm r_j(t_d)$ and goal locations $\bm r_{g_i}(t_d)$, $\bm r_{g_j}(t_d)$.
\item Compare the values of the Lyapunov functions $V^{(k)}(t_d)$ given out of \eqref{sumLyap} for each possible assignment $k\in\{1,2\}$.
\item Move under: $V(t_d)=\min\left\{V^{(k)}(t_d) : \; k\in\{1,2\}\right\}.$
\end{enumerate}
In other words: Without loss of generality, let $k=1$ be the goal assignment before the decision making. The agents implement the following switching logic:
\begin{itemize}
\item If $V^{(1)}(t_d)\leq V^{(2)}(t_d)$, then: $V(t_d)=V^{(1)}(t_d)$, i.e., they keep the same goal assignment,
\item If $V^{(1)}(t_d) > V^{(2)}(t_d)$, then: $V(t_d)=V^{(2)}(t_d)$, i.e., they switch goal assignment.
\end{itemize}
In the sequel, we refer to the decision making based on the logic described above as to the \ac{OGA} policy. Note that in general for $\N>2$ agents, this policy results in locally optimal paths since it re-assigns goals among the connected agents only (which may be $\N_c<\N$ in number), while it does not necessarily apply at the initial time $t_0$, hence to the initial positions of the robots, but rather is triggered at any time $t_d\geq t_0$ that $\N_c\leq \N$ robots may happen to become connected.  

\vspace{3mm}
\begin{Problem}
Establish that the \ac{OGA} policy renders asymptotically stable trajectories for the multi-robot system, i.e., that each agent does converge to some goal location.
\end{Problem}

\subsection{Stability Analysis on the Switched Multi-robot System}
We resort to control analysis tools for switched systems. The closed-loop dynamics of agent $i$ read:
\begin{align}
\label{agent dynamics}
\dot{\bm r}_i(t)=\bm u_i^{(k)}=-\lambda_i\left(\bm r_i(t)-\bm r_{g_i}^{(k)}\right),
\end{align}
while similarly one may write the closed-loop dynamics for the trajectories $\bm r_j(t)$ of agent $j$. The \ac{OGA} policy gives rise to switched dynamics for each agent, in the sense that for $k=1$ each agent moves under assignment matrix $A_1$, while for $k=2$ each agent moves under assignment matrix $A_2$.

Denote $\bm r=\begin{bmatrix}{\bm r_i}^T&{\bm r_j}^T\end{bmatrix}^T$ the state vector of the multi-robot system, which is governed by the switched dynamics:
\begin{align}
\label{switched dynamics}
\dot{\bm r}(t)&=\bm f_k(\bm r(t)), \\
\mbox{where} \quad \bm f_k&=\begin{bmatrix}\bm u_i^{(k)}\\\bm u_j^{(k)}\end{bmatrix}, \; k\in\mathcal K=\{1,2\}.\nonumber
\end{align}
\begin{Assumption}\label{non zeno assumption}
We assume for now that there are only a finite number of switches per unit time.
\end{Assumption}
Denote $n\in\mathbb N$ and consider the sequence of switching times $T=\{t_0, t_1, t_2, t_3, \ldots, t_n,\ldots\}$ and the switching sequence: $\Sigma = \{\bm r_0; (k_0,t_0),(k_1,t_1),\ldots,(k_n,t_n),\ldots\}, \; k_n\in\mathcal K, \forall n\in \mathbb N$. Assuming that the switching sequence is minimal, i.e., that the switching signal does not imply a sequential switching between the \emph{same} subsystems\footnote{Note that this condition is met in our case since in the stability analysis we consider only those switching times when the goal assignment does change.}, it follows that $k_n\neq k_{n+1}$. Without loss of generality, assume that $k_0=1$. Then the (minimal) switching sequence reads: $\Sigma = \{\bm r_0; (1,t_0),(2,t_1),(1,t_3),(2,t_4),\ldots\}$.

\vspace{3mm}
\begin{my switched}\label{my switched system}
The trajectories $\bm r(t)$ of the switched multi-robot system \eqref{switched dynamics} are \ac{gas} \ac{wrt} the goal assignment $k$ under the \ac{OGA} policy. This implies that agents $i$, $j$ converge to their goal locations $\bm r_{g_i}^{(k)}$, $\bm r_{g_j}^{(k)}$.

\begin{proof}
We consider the candidate Lyapunov-like functions $V^{(k)}$, $k\in\{1,2\}$, given out of \eqref{sumLyap}, which encode the motion of the agents under the goal assignment $k$. Out of Lemma \ref{stable} one has that each individual $k$-th subsystem, i.e., the motion of the multi-robot system under the $k$-th assignment, is \ac{ges} i.e., each $V^{(k)}$ is decreasing during the time intervals when the $k$-th subsystem is active. Furthermore, the \ac{OGA} policy dictates that at the switching times $\{t_0, t_1, t_2, t_3, \ldots,\}$ one has:  $$V^{(1)}(t_0)>V^{(2)}(t_1)>V^{(1)}(t_2)>V^{(2)}(t_3)>\ldots,$$ i.e., the value of each Lyapunov-like function $V^{(k)}$ at the beginning of the time intervals when the $k$-th subsystem becomes active satisfies the decreasing condition. This proves that the switched system is stable.

To draw conclusions on asymptotic stability, consider the time intervals $[t_{2n},t_{2n+1})$, $[t_{2n+1},t_{2n+2})$, $n\in\mathbb N$, when the $k=1$ and $k=2$ subsystems are active, respectively. Then the following inequalities hold:
\begin{align*}
V^{(1)}(t_{0}) &> V^{(1)}(t_{1}), \quad \begin{matrix}\mbox{since the subsystem 1 is}\\
\mbox{active on $[t_{0},t_{1})$ and \ac{ges}}\end{matrix}\\
V^{(1)}(t_{1}) &\geq V^{(2)}(t_{1}), \quad \mbox{out of the OGA Policy}\\
V^{(2)}(t_{1}) &>V^{(2)}(t_{2}), \quad \begin{matrix}\mbox{since the subsystem 2 is}\\
\mbox{active on $[t_{1},t_{2})$ and \ac{ges}}\end{matrix}\\
V^{(2)}(t_{2}) &\geq V^{(1)}(t_{2}), \quad \mbox{out of the OGA Policy}\\
V^{(1)}(t_{2}) &> V^{(1)}(t_{3}), \quad \begin{matrix}\mbox{since the subsystem 1 is}\\
\mbox{active on $[t_{2},t_{3})$ and \ac{ges}}\end{matrix}\\
V^{(1)}(t_{3}) &\geq V^{(2)}(t_{3}), \quad \mbox{out of the OGA Policy}\\
&\vdots
\end{align*}
which imply that: $V^{(1)}(t_{2n+2}) < V^{(1)}(t_{2n})$, $\forall n$, i.e.,:
\begin{align*}
V^{(1)}(t_{2n+2}) = \rho_1 V^{(1)}(t_{2n}), \quad 0<\rho_1<1. \;\; \mbox{Then:}
\end{align*}
\begin{align*}
V^{(1)}(t_{2n+2}) - V^{(1)}(t_{2n}) &= - (1-\rho_1)\; \underbrace{V^{(1)}(t_{2n})}_{W(\bm r, \bm r_g^{(1)})},
\end{align*}
where $1-\rho_1>0$, and $W(\bm r, \bm r_g^{(1)})=g\left(\|\bm r-\bm r_g^{(1)}\|^2\right)$ is a continuous, positive definite function \ac{wrt} the goals $\bm r_g^{(1)}$ imposed by the $k=1$ assignment. 
Similarly we obtain: $V^{(2)}(t_{2n+1}) < V^{(2)}(t_{2n+3})$, $\forall n$, which leads to $$V^{(2)}(t_{2n+3}) - V^{(2)}(t_{2n+1}) = - (1-\rho_2)\; \underbrace{V^{(2)}(t_{2n+1})}_{W(\bm r, \bm r_g^{(2)})},$$ 
where $0<\rho_2<1$. Therefore, the switched system is \ac{gas}.
\end{proof}
\end{my switched}

\subsection{Avoiding Zeno Behavior}\label{skdf}

In the sequel we establish that the switching policy does not result in Zeno trajectories \cite{Ames_ACC05}, i.e., in trajectories that converge to some point or switching surface other than the desired equilibrium; such trajectories in practice would result in robots not converging to their respective goal locations.

Given the sequence $T=\{t_0,t_1,\ldots,t_n,\dots\}, \; n\in\mathbb{N},$ of switching times, a switched system is Zeno if there exists some finite time $t_{Z}$ such that:
$$\lim_{n\to\infty}t_n=\sum_{n=0}^\infty(t_{n+1}-t_n)=t_Z.$$ Zeno behavior means that the switching times have a finite accumulation point, i.e., that infinite amount of switchings occurs in a finite time interval. In general, the task of detecting possible Zeno trajectories and extending them beyond their accumulation points is far from trivial \cite{Liberzon} and depends on the problem at hand.

The definition above results in two qualitatively different types of Zeno behavior. A Zeno switched system is \cite{Ames_ACC05}:
\begin{enumerate}
\item[(i)] Chattering Zeno if there exists a finite $C\in\mathbb{N}$ such that $t_{n+1} - t_n = 0$ for all $n > C$.
\item[(ii)] Genuinely Zeno if $t_{n+1} - t_n > 0$ for all $n \in \mathbb N$.
\end{enumerate}
The difference between these is prevalent especially in their detection and elimination. Chattering Zeno behavior results from the existence of a switching surface on which the vector fields ``oppose" each other. This behavior can be eliminated by defining either (i) Filippov solutions which force the flow to ``slide" along the switching surface; this results in a sliding mode behavior, or (ii) hysteresis switching, in order to, not only approximate a sliding mode, but also to maintain the property that two consecutive switching events are always separated by a time interval of positive length.

\vspace{2mm}
The proposed \ac{OGA} policy dictates that a switch among vector fields $\bm f_k$ may occur when $\|\bm r_i(t) - \bm r_j(t)\| \leq R_c$, i.e., when the multi-robot system trajectories $\bm r(t)$ hit the surface $S^c(t) := \left\{\bm r\in \R^{2\N} \;|\; \|\bm r_i(t) - \bm r_j(t)\| = R_c\right\}$. Denote $t_d$ the time instant when the system trajectories lie on the surface $S^c(t_d)$ and the agents are involved in the decision making, $t^{-}_d$, $t^+_d$ the time instants before and after the decision making, respectively, and $k\left(t^-_d\right)$, $k\left(t^+_d\right)$ the assignment before and after the decision making, respectively. The decision making on $S^c(t_d)$ results in two different cases:
\begin{enumerate}
\item The agents decide to keep their goal assignment, i.e., $k\left(t^-_d\right)=k\left(t^+_d\right)$, in which case no switching occurs.
\item The agents decide to swap goals, i.e., $k\left(t^-_d\right)\neq k\left(t^+_d\right)$ and a switching occurs.
\end{enumerate}
\begin{noChatteringZeno}The switched multi-robot system \eqref{switched dynamics} under the \ac{OGA} policy does not suffer from Zeno points.

\begin{proof}\label{nochatteringzeno}
We employ the results in \cite{Ceragioli_NonlinearAnalysis2006}, Theorem 2. Let us assume that the switched multi-robot system \eqref{switched dynamics} has a Zeno point $\bar{\bm r}$. Then it holds that: $\bar{\bm r}\in S^c(t_d)$, $\bar{\bm r}$ is an accumulation point of the set $\mathcal S=\{\bm r\in S^c(t_d):\bm f_k\left(t^-_d\right)=\bm f_k\left(t^+_d\right)\}$, and satisfies: $\nabla S^c(\bar{\bm r}) \bm f_k\left(t^-_d\right)=\nabla S^c(\bar{\bm r}) \bm f_k\left(t^+_d\right)=0$, where $t_d$ is a decision making and switching time instant.

Denote $\bm r_{G_1}=\begin{bmatrix}x_{G_1}&y_{G_1}\end{bmatrix}^T$, $\bm r_{G_2}=\begin{bmatrix}x_{G_2}&y_{G_2}\end{bmatrix}^T$ where $\bm r_{G_1}\neq \bm r_{G_2}$ the position vectors of the goal locations ${G_1}$, ${G_2}$, assigned to the agents $i$, $j$ respectively, at time $t^-_d$. The agents implement the \ac{OGA} policy at time $t_d$ and decide to swap goals. Then, the corresponding vector fields read: $$\bm f_k\left(t^-_d\right)=\begin{bmatrix}-\lambda_i(\bm r_i-\bm r_{G_1})\\-\lambda_j(\bm r_j-\bm r_{G_2})\end{bmatrix}, \; \bm f_k\left(t^+_d\right)=\begin{bmatrix}-\lambda_i(\bm r_i-\bm r_{G_2})\\-\lambda_j(\bm r_j-\bm r_{G_1})\end{bmatrix},$$ while $\nabla S^c(t_d)=\begin{bmatrix}{\bm r_{ij}}^T&-{\bm r_{ij}}^T\end{bmatrix}$, where $\bm r_{ij}=(\bm r_i - \bm r_j)$. It is easy to verify that $\mathcal S=\emptyset$, since $\bm f_k\left(t^-_d\right)=\bm f_k\left(t^+_d\right) \Rightarrow \bm r_{G_1}=\bm r_{G_2}$, a contradiction. Thus, the set of accumulation points of $\mathcal S$ is empty, which implies that no Zeno points $\bar{\bm r}$ are contained there, a contradiction. Thus, the \ac{OGA} policy does not suffer from Zeno points.
\end{proof}
\end{noChatteringZeno}

This result establishes that the decision making under the \ac{OGA} renders trajectories that do not accumulate on, i.e., always escape, the switching surface $S^c(t_d)$; hence the robots do not get stuck away from their assigned goals. Finally, to maintain the property that there always will be a positive time interval $\tau>0$ between consecutive switching times, we implement a hysteresis-like switching logic; the idea is that after the $(i,j)$ decision making at time $t_d$ and while connected, the agents $i$, $j$ do not get involved in a new $(i,j)$ decision making, unless the agents have become disconnected first. Note that in addition, this hysteresis logic validates our Assumption \ref{non zeno assumption} on finite amount of switches per unit time interval.

\subsection{The $\N>2$ Robot Case}
The results on the stability and non-Zeno behavior of the multi-robot system extend to the case of $\N>2$ agents. Consider $\N$ agents that need to converge to $\N$ goal locations. This gives rise to $\N!$ possible goal assignments $k\in\{1,2,\dots,\N!\}$, or in other words, $\N!$ candidate switched subsystems $\bm f_k$ for the switched multi-robot system:
\begin{align}
\label{switched dynamics N}
\dot{\bm r}(t)&=\bm f_k(\bm r(t)), \\
\mbox{where:} \; \bm r&=\begin{bmatrix}{\bm r_1}\\\vdots\\{\bm r_{\N}}\end{bmatrix}, \; \bm f_k=\begin{bmatrix}\bm u_1^{(k)}\\\vdots\\\bm u_{\N}^{(k)}\end{bmatrix}, \; k\in\mathcal K=\{1,\dots,\N!\}\nonumber.
\end{align}

For $\N>2$ agents the decision making on the optimal goal assignment involves the $\N_c\leq \N$ connected agents at time $t_d$. The agents exchange information on their positions and goal locations at time $t_d$, compare the values of the $\N_c!$ Lyapunov functions $V^{(k)}(t_d)$, where $k\in\mathcal K_c(t_d) \subseteq \mathcal K$, and pick the goal assignment $k$ which corresponds to $\min\{V^{(k)}(t_d)\}$.

\subsubsection{Stability Analysis}
Consider the sequence of switching times $T=\{t_0, t_1, \ldots, t_n,\ldots\}$ and the switching sequence $\Sigma = \{\bm r_0; (k_0,t_0),(k_1,t_1),\ldots,(k_n,t_n),\ldots\}$, $k_n\in\mathcal K$. Then, the reasoning and analysis followed in Theorem \ref{my switched system} apply to the $\N>2$ case as well.

\begin{my switched N}
The trajectories $\bm r(t)$ of the switched multi-robot system \eqref{switched dynamics N} are \ac{gas} \ac{wrt} the goal assignment $k$.

\begin{proof}
We consider the candidate Lyapunov-like functions $V^{(k)}$, $k\in\{1,\dots,\N!\}$, encoding the motion of the agents under goal assignment $k$. Out of Lemma \ref{stable}, each individual $k$-th subsystem, i.e., the motion of the multi-robot system under the $k$-th assignment, is \ac{ges}. This implies that each $V^{(k)}$ is decreasing on the time intervals that the $k$-th subsystem is active. Furthermore, the \ac{OGA} policy dictates that at the switching times $\{t_0, t_1, t_2, t_3, \ldots,\}$ one has: $$V^{(k(t_0))}(t_0)>V^{(k(t_1))}(t_1)>V^{(k(t_2))}(t_2)>\ldots,$$ where $k(t_n)$ denotes the $k$-th subsystem that becomes active at time $t_n$, $n\in\mathbb N$; in other words, the value of each Lyapunov-like function $V^{(k)}$ at the beginning of the time intervals when the $k$-th subsystem becomes active satisfies the decreasing condition, i.e., the switched system is Lyapunov stable. To draw conclusions on the asymptotic stability, consider any pair of switching times $t_{n_1}<t_{n_2}$ when the $k$-th subsystem becomes active, the corresponding time intervals $[t_{n_1},t_{m_1})$, $[t_{n_2},t_{m_2})$ while the $k$-th subsystem remains active, with $t_{m_1}<t_{n_2}$, and follow the reasoning as in Theorem \ref{my switched system}, i.e.,:
\begin{align*}
V^{(k)}(t_{n_1}) &> V^{(k)}(t_{m_1}), \quad \begin{matrix}\mbox{since the subsystem $k$ is}\\
\mbox{active on $[t_{n_1},t_{m_1})$ and \ac{ges}}\end{matrix}\\
V^{(k)}(t_{m_1}) &\geq V^{(l)}(t_{m_1}), \quad \mbox{out of the OGA Policy, $l\neq k$}\\
V^{(l)}(t_{m_1}) &\geq V^{(k)}(t_{n_2}), \quad \mbox{out of the OGA Policy}\\
V^{(k)}(t_{n_2}) &> V^{(k)}(t_{m_2}), \begin{matrix}\mbox{since the subsystem $k$ is}\\
\mbox{active on $[t_{n_2},t_{m_2})$ and \ac{ges}}\end{matrix}
\end{align*}
which imply that: $V^{(k)}(t_{m_2}) < V^{(k)}(t_{m_1})$, i.e., that:
\begin{align*}
V^{(k)}(t_{m_2}) = \rho_k V^{(k)}(t_{m_2}), \quad 0<\rho_k<1. \;\; \mbox{Then:}
\end{align*}
\begin{align*}
V^{(k)}(t_{m_2}) - V^{(k)}(t_{m_1}) &= - (1-\rho_k)\; V^{(k)}(t_{m_1}),
\end{align*}
where $1-\rho_k>0$. Following the same reasoning as in Theorem \ref{my switched system}, we have that the switched system is \ac{gas}.
\end{proof}
\end{my switched N}

\subsubsection{Non-Zeno Behavior} Finally, following the same pattern as for the $\N=2$ case, one may verify that:
\begin{noChatteringZenoN}
The switched multi-robot system \eqref{switched dynamics N} under the \ac{OGA} policy does not suffer from Zeno behavior.

\begin{proof}
Assume that the switched system \eqref{switched dynamics N} has a Zeno point $\bm{\bar r}$. Then, $\bm{\bar r}\in S^c(t_d)$ and $\bm{\bar r}$ is an accumulation point of the set $\mathcal S=\{\bm r\in S^c(t_d):\bm f_{k}\left(t^-_d\right)=\bm f_{k}\left(t^+_d\right)\}$, where $S^c(t_d)$ is the switching surface at the decision and switching time $t_d$. The \ac{OGA} policy dictates that at least one pairwise goal swap occurs at time $t_d$, since if the agents decided to keep the goals they had at time $t^-_d$, then $t_d$ would not have been a switching time, a contradiction. Since at least one pair of agents $(i,j)$ switches goal locations, denoted as $\bm r_{G_i}$, $\bm r_{G_j}$, the condition $\bm f_{k}\left(t^-_d\right)=\bm f_{k}\left(t^+_d\right)$ on the switched vector fields holds true only when $\bm r_{G_i}=\bm r_{G_j}$, see the analysis in Theorem \ref{nochatteringzeno}, i.e., a contradiction by construction. Thus, the set $\mathcal S=\emptyset$, which furthermore implies that the set of its accumulation points is empty, implying that no Zeno points can be contained there. Thus, the switched multi-robot system \eqref{switched dynamics N} does not exhibit Zeno behavior.
\end{proof}
\end{noChatteringZenoN}

\section{Collision Avoidance Policy}\label{CA}

The \ac{OGA} policy does not ensure that inter-agent collisions, realized as keeping $d_{ij}(t)\geq d_s$, where $d_s$ is a safety distance, are avoided $\forall t\in[0,\infty)$ (see also \cite{Turpin_WAFR12, Turpin_IJRR14}). 
\vspace{2mm}
\begin{Problem}
Establish (sufficient) conditions under which the \ac{OGA} policy is collision-free.
\end{Problem}

\subsection{Detecting Conflicts}
We are referring to time $t>t_d$, i.e., after $\N_c\leq \N$ connected agents have decided on a goal assignment $k$ based on the \ac{OGA} policy, and move towards their goal locations $\bm r_{g_i}^{(k)}$. In the sequel we drop the notation $\cdot^{(k)}$, in the sense that the goal assignment $k$ is kept fixed.

We would like to identify a metric (a ``supervisor") determining whether the \ac{OGA} policy results in collisions. Let us consider the collision avoidance constraint:
\begin{align}
\label{collision avoidance constraint}
c_{ij}(\bm r_i, \bm r_j)=(x_i-x_j)^2+(y_i-y_j)^2-{d_s}^2>0,
\end{align}
encoding that the inter-agent distance $d_{ij}=\|\bm r_i-\bm r_j\|$ should always remain greater than a safety distance $d_s$.

To facilitate the analysis using Lyapunov-like approaches, we encode the constraint \eqref{collision avoidance constraint} as a Lyapunov-like function. Recall that a barrier function \cite{Nesterov} is a continuous function whose value on a point increases to infinity as the point approaches the boundary of the feasible region; therefore, a barrier function is used as a penalizing term for violations of constraints. The concept of \ac{rbf} in particular was introduced in \cite{Wills_ACC02} in order to not only regulate the solution to lie in the interior of the constrained set, but also to ensure that, if the system converges, then it converges to a desired point.

We define the logarithmic barrier function $b_{ij}(\cdot):\R^2\times\R^2\to\R$ for the constraint \eqref{collision avoidance constraint} as:
\begin{align}
\label{barrier}
b_{ij}(\bm r_i, \bm r_j)=-\ln\left({c_{ij}(\bm r_i, \bm r_j)}\right),
\end{align}
which tends to $+\infty$ as $c_{ij}(\cdot)\rightarrow 0$, i.e., as $d_{ij}\to d_s$. The \ac{rbf} of \eqref{barrier} is defined as \cite{Wills_ACC02}:
\begin{align}
\label{recentered barrier function}
r_{ij} = b_{ij}(\bm r_i,\bm r_j) - b_{ij}(\bm r_{g_i},\bm r_j)-{\nabla b_{ij}}^T\big|_{\bm r_{g_i}} (\bm r_i - \bm r_{g_i}),
\end{align}
where $\bm r_{g_i}=\begin{bmatrix}x_{g_i}&y_{g_i}\end{bmatrix}^T$ is the goal location for agent $i$, $b_{ij}(\bm r_{g_i},\bm r_j)$ is the value of \eqref{barrier} evaluated at $\bm r_{g_i}$, $\nabla b_{ij}=\begin{bmatrix}\frac{\partial b_{ij}}{\partial x_i}&\frac{\partial b_{ij}}{\partial y_i}\end{bmatrix}^T$ is the gradient vector of the function $b_{ij}(\cdot)$, and ${\nabla b_{ij}}^T\big|_{\bm r_{g_i}}$ is the transpose of the gradient vector, evaluated at $\bm r_{g_i}$. 
\vspace{2mm}
To ensure that we have a nonnegative function to be used as a Lyapunov-like function, we define \cite{Panagou_TAC14}:
\begin{align}
\label{squared recentered collision avoidance}
w_{ij}(\cdot)=\left({r_{ij}(\bm r_i, \bm r_j, \bm r_{g_i})}\right)^{2},
\end{align}
which now is a positive definite function. To furthermore encode that the position trajectories of agent $i$ remain bounded in a prescribed region (for reasons that will be explained in the technical analysis later on), for each agent $i$ we define the workspace constraint:
\begin{align}
\label{boundary constraint}
c_{i0}(\bm r_i, \bm r_0)=R_0^2-(x_i-x_0)^2-(y_i-y_0)^2>0,
\end{align}
which encodes that the position $\bm r_i=\begin{bmatrix}x_i&y_i\end{bmatrix}^T$ of agent $i$ should always lie in the interior of a circle of center $\bm r_0=\begin{bmatrix}x_0&y_0\end{bmatrix}^T$ and radius $R_0>R_c$. This can be thought as the workspace where the agents operate. Then, in the same reasoning followed before, we define the barrier function:
\begin{align}
b_{i0}(\bm r_i, \bm r_0)=-\ln\left({c_{i0}(\bm r_i, \bm r_0)}\right),
\end{align}
and its corresponding \ac{rbf}:
\begin{equation}
r_{i0} = b_{i0}(\bm r_i,\bm r_0) - b_{i0}(\bm r_{g_i},\bm r_0)-{\nabla b_{i0}}^T\big|_{\bm r_{g_i}} (\bm r_i - \bm r_{g_i}),
\end{equation}
which vanishes only at the goal location $\bm r_{g_i}$ and tends to infinity as the position $\bm r_i$ tends to the workspace boundary. To get a positive definite function we consider:
\begin{align}
w_{i0}(\cdot)=\left({r_{i0}(\bm r_i, \bm r_0, \bm r_{g_i})}\right)^{2}.
\end{align}
Therefore, an encoding that agent $i$ stays $d_s$ away from all its neighbor agents $j$, while staying within the bounded workspace, can be now given by an approximation of the maximum function of the form:
\begin{align}
\label{approximation}
w_i=\left((w_{i0})^{\delta}+\sum\limits_{j\in\mathcal N_i} (w_{ij})^{\delta}\right)^{\frac{1}{\delta}},
\end{align}
where $\delta\in[1,\infty)$, and $\mathcal N_i \subseteq \mathcal N$ is the set of connected (neighbor) agents $j\neq i$ of agent $i$. The function $w_i$ vanishes at the goal location $\bm r_{g_i}$ and tends to $+\infty$ as at least one of the terms $w_{i0}$, $w_{ij}$ tends to $+\infty$, i.e., as at least one of the inter-agent distances $d_{ij}$ tends to $d_s$, or as agent $i$ tends to the workspace boundary. Finally, to ensure that we have a Lyapunov-like function for agent $i$ which uniformly attains its maximum value on constraints' boundaries we take:
\begin{align}
\label{Lyapunov-like}
W_i = \frac{w_i}{1 + w_i},
\end{align}
which is zero for $w_i = 0$, i.e., at the goal position $\bm r_{g_i}$ of agent $i$, and equal to 1 as $w_i\to\infty$. For more details on the analytical construction the reader is referred to \cite{Panagou_TAC14, Stipanovic_MonotoneApproximations2012}.

The level sets of the Lyapunov-like function \eqref{Lyapunov-like} can now be used as a sufficient criterion on determining whether the control inputs out of the \ac{OGA} policy jeopardize safety. Let us consider the time derivative of \eqref{Lyapunov-like}:
\begin{align}
\label{dotVj}
\dot W_i &= \begin{bmatrix}\frac{\partial W_i}{\partial x_i}&\frac{\partial W_i}{\partial y_i}\end{bmatrix}\begin{bmatrix}\dot x_i\\\dot y_i\end{bmatrix}+\sum_{j\in\mathcal N_i}\left(\begin{bmatrix}\frac{\partial W_i}{\partial x_j}&\frac{\partial W_i}{\partial y_j}\end{bmatrix}\begin{bmatrix}\dot x_j\\\dot y_j\end{bmatrix}\right) = \nonumber\\
&={\bm \zeta_{i}}^T \bm u_i+\sum_{j\in\mathcal N_i}\left({\bm \zeta_{ij}}^T \bm u_j\right),
\end{align}
where $\bm \zeta_i\triangleq\begin{bmatrix}\frac{\partial W_i}{\partial x_i}&\frac{\partial W_i}{\partial y_i}\end{bmatrix}^T$, $\bm \zeta_{ij}\triangleq\begin{bmatrix}\frac{\partial W_i}{\partial x_j}&\frac{\partial W_i}{\partial y_j}\end{bmatrix}^T$. This time derivative is defined almost everywhere, except for a set of measure zero corresponding to the case when the gradient term $\nabla b_{ij}|_{\bm r_{g_i}}$ is undefined (namely, this corresponds to $\bm r_j = \bm r_{g_i}$). Not surprisingly, the evolution of the time derivative \eqref{dotVj} along the trajectories of agent $i$ depends not only on its own motion (through $\bm u_i$), but also on the motion of its neighbor agents $j\neq i$ through their velocities $\bm u_j$. This time derivative provides sufficient conditions on establishing safety under the \ac{OGA} policy. To see how, let us consider the following Lemma.
\begin{barrier condition}\label{barrier condition}
If the control inputs $\bm u_i$, $\bm u_j$, where $j\in\mathcal N_i$, out of the \ac{OGA} policy satisfy the following condition for all $i\in\{1,\dots,\N\}$:
\begin{align}
\label{safe condition}
{\bm \zeta_{i}}^T\bm u_i+\sum_{j\in\mathcal N_i}\left({\bm \zeta_{ij}}^T\bm u_j\right)< 0,
\end{align}
then the \ac{OGA} policy is asymptotically stable \ac{wrt} the current goal assignment $k$ and furthermore collision-free.

\begin{proof}
To verify the argument, consider the properties of the Lyapunov-like function \eqref{Lyapunov-like} and denote the constrained set for each agent $i$ as $\mathcal K_i=\{\bm r\in \R^{2\N} \;|\; c_{ij}(\cdot)\geq 0\}$, where $j\in\mathcal N_i\cup \{0\}$. The set $\mathcal K_i$ is by construction compact (i.e., closed and bounded), with the level sets of $W_i$ being closed curves contained in the set $\mathcal K_i$. Then, the condition \eqref{safe condition} implies that the system trajectories $\bm r_i(t)$ under the \ac{OGA} control input $\bm u_i$ evolve along the negative direction of the gradient vector $\nabla W_i=\begin{bmatrix}\frac{\partial W_i}{\partial x_i}&\frac{\partial W_i}{\partial y_i}\end{bmatrix}$ of $W_i$, i.e., always remain in the interior of the constrained set $\mathcal K_i$; this further reads that the inter-agent distances $d_{ij}(t)$ never become smaller than $d_s$.
\end{proof}
\end{barrier condition}

\vspace{2mm}
\begin{Problem}
Seek a policy that becomes active when the \ac{OGA} policy is about to result in colliding trajectories, and ensures that the inter-agent distances $d_{ij}$ remain greater than the critical distance $d_s$.
\end{Problem}

\vspace{1mm}

We develop the \ac{CA} policy described in the following section.

\subsection{Resolving Conflicts}
Let us consider the switching surface: 
\begin{equation}
\mathcal Q_i={\bm \zeta_{i}}^T\bm u_i+\sum_{j\in\mathcal N_i}\left({\bm \zeta_{ij}}^T\bm u_j\right).
\end{equation}
Then for $\mathcal Q_i<0$ one has out of Lemma \ref{barrier condition} that the \ac{OGA} policy is both \ac{gas} and collision-free. For $\mathcal Q_i>0$ one has $\dot W_i>0$, which implies that the position trajectories $\bm r_i(t)$ evolve along the positive direction of the gradient vector $\nabla W_i$. This condition \emph{may} jeopardize safety and dictates the definition of an additional policy to ensure that the trajectories $\bm r_i(t)$ remain in the constrained set $\mathcal K_i$.

At this point, one may be tempted to use a \ac{CA} control law for agent $i$ along the gradient-descent direction of $W_i$, to render the trajectories of the multi-robot system collision-free in the set $\mathcal Q_i>0$. However, such control laws typically grow unbounded for system trajectories approaching local minimum points, i.e., as $\|\nabla W_i\|\to 0$. This may lead to implementation issues when it comes to realistic robotic agents and therefore is undesirable from a practical point-of-view. 

We thus emulate the gradient descent performance along the level sets of \eqref{Lyapunov-like} with control inputs that are always well-defined. We define a vector field $\mathbf F_i: \R^2 \rightarrow \R^2$ to be used as a feedback motion plan for each agent $i$, in the spirit developed in our earlier work \cite{Panagou_submitted}. The analytical expression is given as:
\begin{align}
\label{feedback motion plan dynamic}
\mathbf F_i = \prod_{j\in\mathcal{N_i}} (1-\sigma_{ij}) \mathbf F^i_{g} + \sum_{j\in\mathcal{N_i}} \sigma_{ij} \mathbf F^i_{oj},
\end{align}
where:
\begin{align}
\label{bump function}
\sigma_{ij} = \left\{
              \begin{array}{ll}
                1, & \; \hbox{for }\;d_s \leq d_{ij}\leq R_s; \\
                a\;{d_{ij}}^3 + b\;{d_{ij}}^2 + c\; d_{ij} + d, & \; \hbox{for }\;R_s< d_{ij}< R_c; \\
                0, & \; \hbox{for }\;d_{ij}\geq R_c;
              \end{array}
            \right.
\end{align}
the attractive to the goal position $\bm r_{g_i}=\begin{bmatrix}x_{g_i}&y_{g_i}\end{bmatrix}^T$ vector field $\mathbf F^i_{g}$ is defined as: 
\begin{subequations}
\begin{align}
\F^i_{xg} &= -\frac{x-x_{g_i}}{\sqrt{(x-x_{g_i})^2+(y-y_{g_i})^2}}, \\
\F^i_{yg} &= -\frac{y-y_{g_i}}{\sqrt{(x-x_{g_i})^2+(y-y_{g_i})^2}},
\end{align}
\end{subequations}
the repulsive vector field $\mathbf F^i_{oj}$ around each agent $j\neq i$ is defined as:
\begin{subequations}
\label{normalized repelling node}
\begin{align}
\F^i_{xoj}= \frac{x_i-x_j}{\sqrt{(x_i-x_j)^2+(y_i-y_j)^2}},\\
\F^i_{yoj}= \frac{y_i-y_j}{\sqrt{(x_i-x_j)^2+(y_i-y_j)^2}},
\end{align}
\end{subequations}
$R_s$ the radius of the circle where the repulsive vector field is active, the coefficients $a$, $b$, $c$ and $d$ have been computed as:
\begin{align*}
a &= -\frac{2}{(R_s-R_c)^3}, & b &= \frac{3(R_s+R_c)}{(R_s-R_c)^3},\\
c &= -\frac{6\;R_s R_c}{(R_s-R_c)^3}, & d &= \frac{{R_c}^2(3R_c-R_s)}{(R_s-R_c)^3},
\end{align*}
so that \eqref{bump function} is a $\mathcal C^1$ function. The vector field \eqref{feedback motion plan dynamic} for each agent $i$ has integral curves that are:
\begin{enumerate} 
\item convergent to the goal location $\bm r_{g_i}$ as long as $d_{ij}>R_c$ for all $j\neq i$, i.e., when no agent $j$ lies in the communication region $\mathcal C_i$ of agent $i$, and 
\item repulsive \ac{wrt} agents $j$ which lie within distance $d_{ij}<R_s$ \ac{wrt} agent $i$, i.e., within the communication region $\mathcal C_i$ of agent $i$. 
\end{enumerate} 
Most importantly, the integral curves of \eqref{feedback motion plan dynamic} are always well-defined within the constrained set $\mathcal K_i$ of agent $i$, in contrast to the gradient vector field out of the Lyapunov-like barrier function \eqref{Lyapunov-like} that is undefined at the points where the function \eqref{Lyapunov-like} is not defined. The vector field \eqref{feedback motion plan dynamic} by construction vanishes only at the goal location $\bm r_{g_i}$. Thus, the vector field \eqref{feedback motion plan dynamic} dictates a safe reference direction $\phi_i=\atan2(\F_y^i, \F_x^i)$ everywhere on $\mathcal K_i\setminus \{\bm r_{g_i}\}$, where $\F_x^i$, $\F_y^i$ are the components of \eqref{feedback motion plan dynamic}. Furthermore, one can show (\cite{Panagou_submitted}, Theorem 5) that the vector field \eqref{feedback motion plan dynamic} is an almost globally convergent feedback motion plan for agent $i$, i.e., all agent trajectories along the integral curves of \eqref{feedback motion plan dynamic} converge to the goal location $\bm r_{g_i}$, except for those starting from a set of initial conditions of measure zero. For more details, the reader is referred to \cite{Panagou_submitted}. 

Now, in the ideal case of a static environment, i.e., when agents $j\neq i$ are not moving, the control policy on the linear speed $u_i$ of agent $i$ can naturally be independent of the linear speeds $u_j$ of other agents $j\neq i$, in the sense that these are identically zero. In the practical case of dynamic environments however, i.e., when agents $j\neq i$ are moving, the control policy on the magnitude $u_i$ of the linear velocity vector $\bm u_i$ of each agent $i$ should take into account the motion of neighbor agents as well, so that the inter-agent distances $d_{ij}(t)$ are guaranteed to remain greater than the lower bound $d_s$, $\forall t\geq 0$. 

We use the following velocity coordination protocol to govern the speed $u_i$ for each agent $i$:  
\begin{last resort}\label{last resort}
Let $\N_c\leq \N$ agents become connected at some time instant $t_d\geq 0$ while moving under some goal assignment $k\in\{1,\dots,\N_c!\}$. The motion of each agent $i$ in the connected group is collision-free under the control law: 
\begin{align}
\label{uj}
\bm u_{i}= \left(\min\limits_{j\in \mathcal N_i | J_i<0} \{u_{i|j}\}\right) \; \bm{\eta}_i,
\end{align}
where: $\bm{\eta}_i=\begin{bmatrix}\cos\phi_i & \sin\phi_i\end{bmatrix}^T$ is the unit vector along the direction of the vector field \eqref{feedback motion plan dynamic}, $\bm r_{ij}=\bm r_i-\bm r_j$, $ J_i={\bm r_{ij}}^T\bm u_i$, $u_{is|j} = \frac{{\bm r_{ij}}^T\bm u_j}{{\bm r_{ij}}^T\bm \eta_i}$, $0<\varepsilon_i<1$, and
\begin{align}
u_{i|j} &= u_{i}\; \frac{d_{ij}-d_s}{R_c-d_s}+\varepsilon_i \underbrace{\frac{{\bm r_{ij}}^T\bm u_j}{{\bm r_{ij}}^T\bm \eta_i}}_{u_{is|j}}\frac{R_c-d_{ij}}{R_c-d_s}.\label{safe vel}
\end{align}
\end{last resort}

\begin{Remark}
In effect, the control law \eqref{uj} forces agent $i$ to adjust its speed \ac{wrt} the worst-case neighbor $j\in \mathcal N_i$, i.e., \ac{wrt} the neighbor $j$ towards whom the rate of change of inter-agent distance $d_{ij}$ is maximized via the motion of agent $i$. 
\end{Remark}

\begin{proof}
Our goal is to show that the constrained set $\mathcal K_i$ encoded via the outermost level set of the Lyapunov-like barrier \eqref{Lyapunov-like} is rendered a positively invariant set under the control law \eqref{uj}. We employ the fundamental characterization of positively invariant sets \cite{Blanchini_Miani} through the necessary and sufficient conditions of Nagumo's theorem on the boundary of the constrained set $\mathcal K_i$ of each agent $i$, namely: $\frac{d}{dt}\; c_{ij}(\bm r_i, \bm r_j)\geq 0, \; \forall \bm r_j \in\partial \mathcal K_{i}, \; \forall (i,j)$.

Consider that at some time $t_d\geq 0$, the distance $d_{ij}(t_d)$ between a pair of connected agents $(i,j)$ is $d_{ij}(t_d)\leq R_c$. Consider the time derivative of the inter-agent distance, which after some calculations reads:
\begin{align}
\label{dijdt}
\frac{d}{dt}\;d_{ij}&\overset{\eqref{system}}{=}\frac{{\bm r_{ij}}^T (\bm u_i - \bm u_j)}{d_{ij}}.
\end{align} 

\begin{figure}
\centering
\includegraphics[width=0.8\columnwidth,clip]{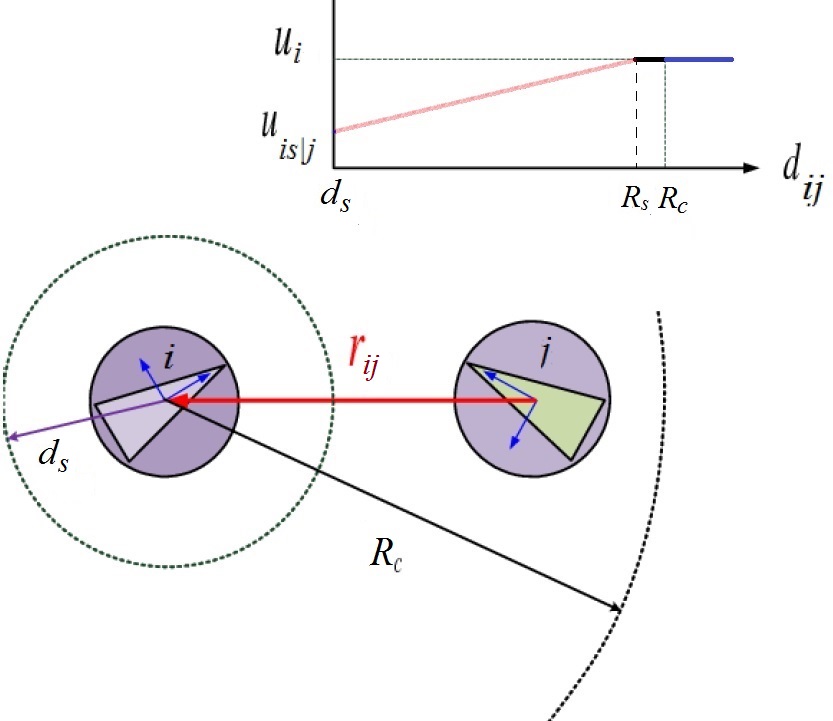}
\caption{If $J_{j}\triangleq{\bm r_{ij}}^T\bm \eta_i < 0$, i.e., if agent $i$ moves towards agent $j$, then agent $i$ adjusts its linear speed according to the velocity profile shown here, given analytically by \eqref{uj}.}
\label{fig:velocity adjustment}
\end{figure}
Under the control law \eqref{uj}, agent $i$ adjusts its linear speed $u_i$ according to the velocity profile shown in Fig. \ref{fig:velocity adjustment}, given analytically out of \eqref{uj}, so that the distance $d_{ij}$ \ac{wrt} the worst case neighbor $j$ remains greater than $d_s$. The worst case neighbor is the agent $j\in \{\mathcal N_{i} \; | \; {J_i<0}\}$ towards whom the rate of change of relative distance $d_{ij}$ given by \eqref{dijdt}, due to the motion of agent $i$, is maximum. More specifically: The term $j\in \{\mathcal N_i \; | \; J_i<0\}$ describes the set of neighbor agents $j$ of agent $i$ towards whom agent $i$ is moving \cite{Panagou_TAC14}. Agent $i$ computes safe velocities $u_{i|j}$ given out of \eqref{safe vel} \ac{wrt} each neighbor in this set, and picks the minimum $\min\{u_{i|j}\}$ of the safe velocities $u_{i|j}$ so that the first term in \eqref{dijdt} is as small as possible. Now, the value of the safe velocity $u_{i|j}$ \eqref{safe vel} when $d_{ij}=d_s$ is equal to $\varepsilon_i u_{is|j}=\varepsilon_i\frac{{\bm r_{ij}}^T\bm u_j}{{\bm r_{ij}}^T\bm \eta_i}$. Plugging this value into \eqref{dijdt} reads: 
\begin{align*}
\dot d_{ij}\big |_{d_{ij}=d_s}=\frac{(\varepsilon_i-1) u_j \; {\bm r_{ij}}^T\bm \eta_j}{d_{ij}}\geq 0.  
\end{align*}
To see why this condition is true, recall that $\varepsilon_i-1<0$, $u_j>0$, and ${\bm r_{ij}}^T\bm \eta_j\leq 0$: this is because agent $j$ is either following a vector field $\mathbf F_j$ that points away from agent $i$, or happens to move away from agent $i$ in the first place. This implies that the inter-agent distance $d_{ij}$ can not become less than $d_s$, i.e., that collisions are avoided. 
\end{proof}

\subsection{Ensuring both Stability and Safety}
We now define the switching policy that realizes the collision-free motion for each agent $i$ as follows:

\vspace{2mm} 

Let $\N_c\leq \N$ agents become connected at some time instant $t_d\geq 0$ while moving under some goal assignment $k\in\{1,\dots,\N_c!\}$. Define $$\mathcal R_i(t_d)=\cos(\theta_i(t_d)) \cos(\phi_i(t_d))$$ the decision-making surface for agent $i$, where $\theta_i(t_d)$ is the direction of the velocity vector $\bm u_i$ at time $t_d$ given out of the \ac{OGA} controller \eqref{controller}, and $\phi_i(t_d)$ is the direction of the vector field vector $\mathbf F_i$ at time $t_d$, given out of \eqref{feedback motion plan dynamic}. Then:
\begin{itemize}
\item If $\mathcal R_i(t_d)>0$, the trajectories under the \ac{OGA} policy are safe, and agent $i$ keeps moving under \eqref{controller}. 
\item If $\mathcal R_i(t_d)\leq 0$, the trajectories out of the \ac{OGA} policy might violate safety, and agent $i$ moves under \eqref{uj}.
\end{itemize}

The switching between the \ac{OGA} policy and the \ac{CA} policy across the switching surface $\mathcal R_i=0$ gives rise to closed-loop dynamics with discontinuous right-hand side for each agent $i$:
\begin{align}
\label{switched-oga-ca}
\dot{\bm r}_i(t)=\bm v_i^{(p)},
\end{align}
where $p\in\mathcal P=\{1,2\}$, and $\bm v_i^{(p)}$ is the closed-loop dynamics of agent $i$ under the \ac{OGA} policy \eqref{controller} and the \ac{CA} policy \eqref{uj}, respectively. More specifically, for $p=1$, agent $i$ moves under the \ac{OGA} policy, while for $p=2$, agent $i$ moves under the \ac{CA} policy. We formalize the switching logic between the resulting closed-loop subsystems via a hysteresis technique, similar to the one used in Section \ref{skdf}; i.e., after making the decision between either subsystem $p=1$ or subsystem $p=2$, the agents do not get involved in a new decision making unless they become disconnected first. The hysteresis logic gives rise to a switching sequence of times $\mathcal T=\{\tau_1, \tau_2, \tau_3, \dots, \tau_n, \dots, \}$, where the time interval $\tau_{n}-\tau_{n-1}$ is finite $\forall n\in\mathbb N$, and to the switching sequence:
\begin{align*}
\Sigma^\star = \{\bm r_i(\tau_0); (1,\tau_0),(2,\tau_1),\ldots,(1,\tau_{2q}),(2,\tau_{2q+1}),\ldots,\},
\end{align*}
where $\tau_0>t_d$, $q\in\mathbb N$, $\tau_{2q}$ are the time instants when the subsystem $p=1$ (\ac{OGA} policy) is ``switched on", $\tau_{2q+1}$ are the time instants when the subsystem $p=2$ (\ac{CA} policy) is ``switched on".

\begin{Problem}
Establish that the proposed switching strategy $\Sigma^\star$ between the \ac{OGA} policy and the \ac{CA} policy renders the multi-robot trajectories asymptotically stable \ac{wrt} to the assigned goals, and also collision-free.
\end{Problem}

For the stability analysis of this switching strategy, it is inefficient to employ Branicky's decreasing condition on the ``switched-on" time intervals of each individual subsystem; the main reason is that in principle the \ac{CA} strategy might result in trajectories that either are not monotonically decreasing, or are even increasing with a bounded growth for a finite amount of switches. Therefore we resort to results on the stability of switched systems that remove this condition in order to draw conclusions on asymptotic stability.

\vspace{2mm}

\begin{my switched}
The trajectories of the switched multi-robot system under the switching logic $\Sigma^\star$ are \begin{inparaenum}\item[(i)] collision-free, and \item[(ii)] (almost globally)\footnote{The almost global characterization is added for completeness due to the set-measure initial conditions from which at least one of the agents is forced to their respective goal under the \ac{CA} policy.} asymptotically stable \ac{wrt} the assigned goals.\end{inparaenum}

\begin{proof}
The first argument is proved in Lemmas \ref{barrier condition} and \ref{last resort}. The second argument can be verified by directly applying Theorem \ref{zhao}. To this end, let us first note that Lemma \ref{stable} implies that the function $V_i$ for each agent $i$ serves as generalized Lyapunov-like function for the subsystem $p=1$. Furthermore, $W_i$ is a valid generalized Lyapunov-like function almost everywhere for each agent $i$ for the subsystem $p=2$. To see why, consider that the vector field \eqref{feedback motion plan dynamic} is by construction transverse to the boundary of the constrained set $\mathcal K_i$, and convergent to the goal location $\bm r_{g_i}$, except for a set of initial conditions of measure zero \cite{Panagou_submitted}. Hence, during the time intervals when the subsystem $p=2$ is active, the growth, if any, of the function $W_i$ along the system trajectory is bounded. Furthermore, the stability condition \eqref{zhao condition} for each subsystem $p$ is satisfied out of the boundedness of the solutions $\bm r_i(t)$ within the constrained set $\mathcal K_i$, which is by construction closed and bounded. Therefore, the switched system is stable. 

To establish (almost global) asymptotic stability, it should hold that for at least one of the individual subsystems $p$, the value of the corresponding generalized Lyapunov-like function decreases along the sequence of switching times. Let us assume that this is not the case; then this would imply that the closed-loop switched vector fields $\bm v_{i}^{(p)}$, $p\in\{1,2\}$, oppose each other and cancel out on the switching surface $\mathcal R_i$, forcing the system trajectories to get stuck there. In other words, this would imply that the switching policy $\Sigma^\star$ exhibits Zeno trajectories. A Zeno point ${\bar{\bm r}}\in S^c(t_d)$ would be an accumulation point of the set $\mathcal S=\{\bm r\in S^c(t_d):\bm v^{(p)}\left(t^-_d\right)=\bm v^{(p)}\left(t^+_d\right)\}$, and satisfies: $\nabla S^c({\bar{\bm r}}) \bm v^{(p)}\left(t^-_d\right)=\nabla S^c(\bar{\bm r}) \bm v^{(p)}\left(t^+_d\right)=0$, where $t_d$ is a decision making and switching time instant. For each agent $i$, the first condition reads: 
\begin{equation}
\label{cond1}
-\lambda_i(\bar{\bm r}_i-\bm r_{g_i})=u_{i|j}(\bar{\bm r}_i, \bar{\bm r}_j)\;\bm\eta_i(\bar{\bm r}_i, \bar{\bm r}_j), \quad j\in\mathcal N_i,
\end{equation}
where the left-hand side is given out of \eqref{controller} and the right-hand side is given out of \eqref{uj}, with $\bm{\eta}_i(\bar{\bm r}_i, \bar{\bm r}_j)$ the unit vector along the direction $\phi_i$ of the vector field \eqref{feedback motion plan dynamic}; recall that $\phi_i$ depends by construction on the states of the neighbor agents $j\in\mathcal N_i$, hence the dependence on $\bar{\bm r}_j$ in the right-hand side of \eqref{cond1}. Geometrically, the condition \eqref{cond1} implies that the candidate set $\mathcal S$ of Zeno points lies on the line that connects the position of the agent with its goal location and is concurrently collinear with the vector field \eqref{feedback motion plan dynamic}. The second set of conditions reads: 
\begin{subequations}
\begin{align}
-\lambda_i \nabla \left(\bar{\bm r}_i-\bm r_{g_i}\right)&=0, \\
\nabla\left(u_{i|j}(\bar{\bm r}_i, \bar{\bm r}_j)\bm \eta_i(\bar{\bm r}_i, \bar{\bm r}_j)\right)&=0.
\end{align}
\end{subequations} 
The former one reduces to $\lambda_i=0$, $\forall i$, which is a contradiction. Hence the candidate set of Zeno points is empty, i.e., the switching strategy between the \ac{OGA} policy and the \ac{CA} policy does not exhibit infinite switching between the \ac{OGA} and the \ac{CA} policies around an accumulation point. Therefore, the switched multi-robot system under the switching strategy $\Sigma^\star$ does not exhibit Zeno trajectories, and is almost globally asymptotically stable.
\end{proof}
\end{my switched}

\subsection{Robustness against Communication Failures}
We note that establishing robustness against uncertainty (e.g., against sensor noise, communication dropouts and/or delays) is beyond the scope of the current paper. As a special case, we only consider how the case of failed communication can be handled within the centralized version of the proposed approach. Note that finding the set of initial conditions that result in collision-free trajectories for a given distribution of goals and for every possible decentralized assignment is essentially a backwards reachability problem, which is intractable for even low-dimensional systems, i.e., for teams of no more than 2-3 robots. However, the centralized case as treated in the earlier work of the authors \cite{Turpin_WAFR12} offers a (conservative, yet guaranteed) robustness measure for the case of failed communication in a decentralized setting; more specifically, we have the following corollary: 
\begin{Corollary}
Let the multi-robot system under switched agent dynamics given by \eqref{switched-oga-ca}, and the switching sequence $\Sigma^\star$. If for all $\bm r_i(0)$, $\bm r_j(0)$, $\bm r_{G_i}$, $\bm r_{G_j}$, $i,j\in\{1,\dots, N\}$, it holds that $\|\bm r_i(0)-\bm r_j(0)\|\geq 2R\sqrt{2}$, and $\|\bm r_{G_i}-\bm r_{G_j}\|\geq 2R\sqrt{2}$, where $R$ is the radius of the robots, then under failed communication the multi-robot trajectories are collision-free.
\end{Corollary}
\begin{proof}
The corollary states that if the initial robot positions and goal locations are $2R\sqrt{2}$ apart, then the resulting multi-robot trajectories are collision-free, even when the agents' communication fail and no goal swap or collision avoidance controllers are implemented. Note that under failed communication, the affected agents will not be involved in any decision on re-assigning their goals (i.e., will keep their current goals), and will not update their control laws. Hence, whether the trajectories will be collision-free depends on the initial conditions (positions) of the robots and the spatial distribution of the goal locations. The analysis of the centralized version of the \ac{OGA} policy in \cite{Turpin_WAFR12} proves that under the stated assumptions, the \ac{OGA} policy results in collision-free trajectories. This completes the proof.  
\end{proof}

\section{Simulation Results}\label{Simulation Results}

\subsection{Simulations for $\N=15$ and $\N=40$ agents}

Simulation results are provided to evaluate the performance of the switched multi-robot system under the proposed switching logic and control policies. In the interest of space, and given that the representation of large groups of robots results in rather congested figures, here we consider only two scenarios involving $\N=15$ and $\N=40$ agents, respectively. 

\begin{figure}
{
\centering
\subfigure[Initial positions of the agents and the straight paths to the goals determined by the random initial goal assignment.]{
\includegraphics[width=0.85\columnwidth,clip]{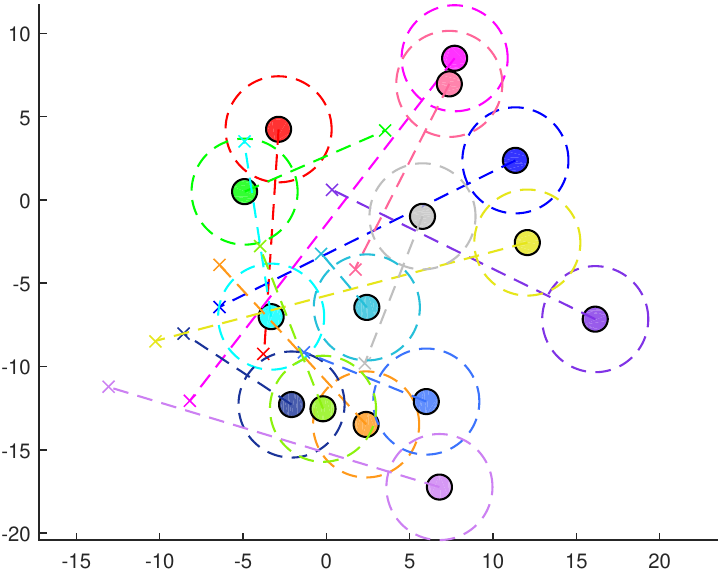}\label{fig:init15}
}
\subfigure[The actual paths to the goals under the proposed control strategy.]{
\includegraphics[width=0.85\columnwidth,clip]{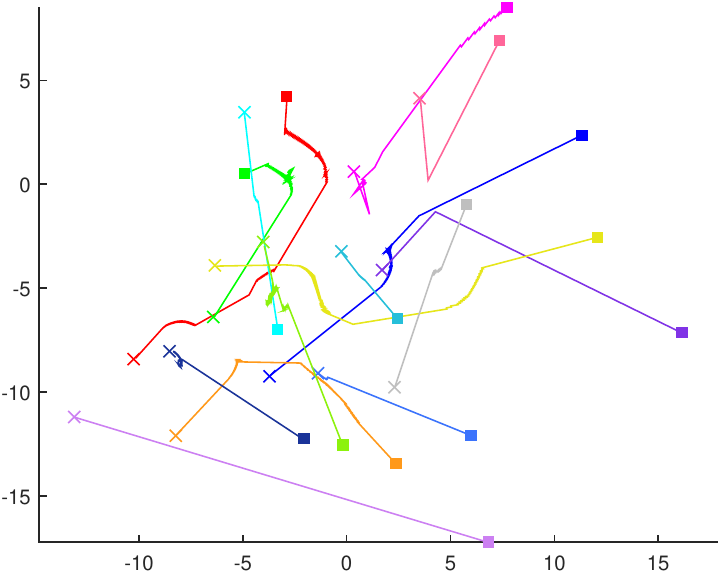}\label{fig:paths15}
}
\subfigure[The final locations of the agents.]{
\includegraphics[width=0.85\columnwidth,clip]{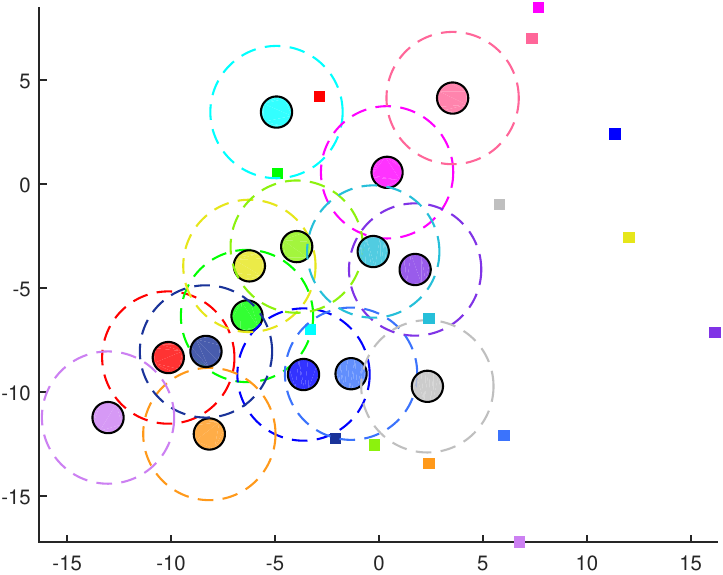}\label{fig:final15}
}
}
\caption{A simulation with $N=15$ robots. The paths followed are straight when in the \ac{OGA} segments, but are potentially curved when using the \ac{CA} control law.}
\label{fig:15RobotSim}
\end{figure}

\begin{figure}
{
\centering
\subfigure[Initial positions of the agents and the straight paths to the goals determined by the random initial goal assignment.]{
\includegraphics[width=0.8\columnwidth,clip]{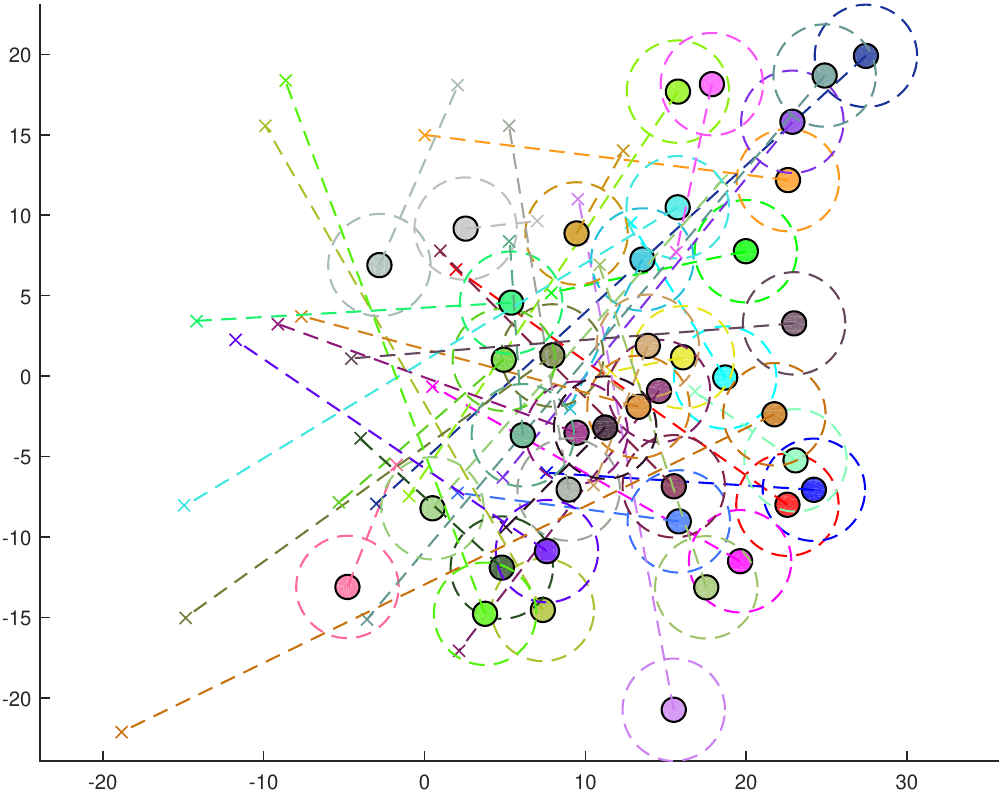}\label{fig:init40}
}
\subfigure[The actual paths to the goals under the proposed control strategy.]{
\includegraphics[width=0.8\columnwidth,clip]{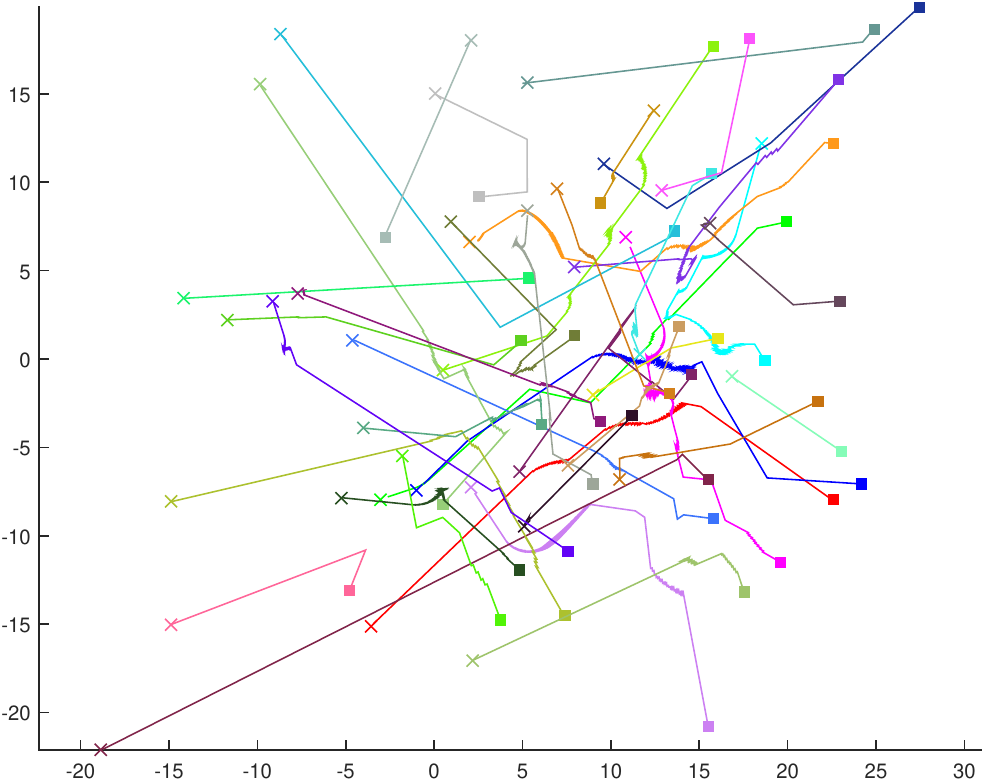}\label{fig:paths40}
}
\subfigure[The final locations of the agents.]{
\includegraphics[width=0.81\columnwidth,clip]{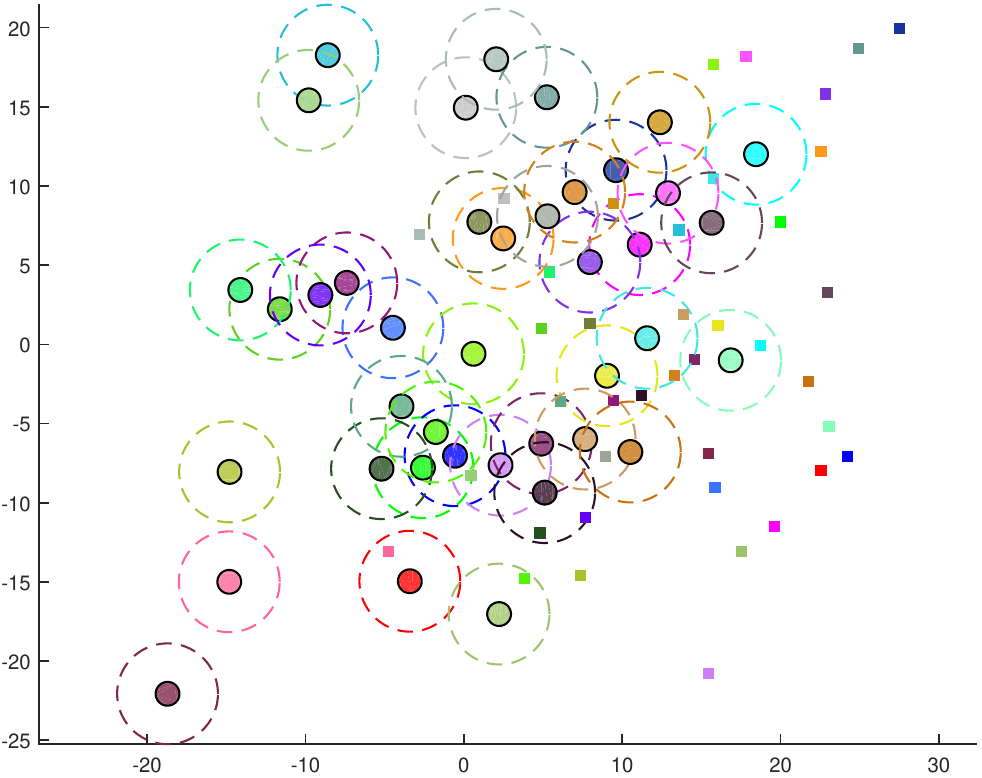}\label{fig:final40}
}
}
\caption{A simulation with $N=40$ robots. The paths are straight during the \ac{OGA} segments, but are potentially curved when using the \ac{CA} control law.}
\label{fig:40RobotSim}
\end{figure}

\begin{figure}
\centering
\includegraphics[width=0.975\columnwidth,clip]{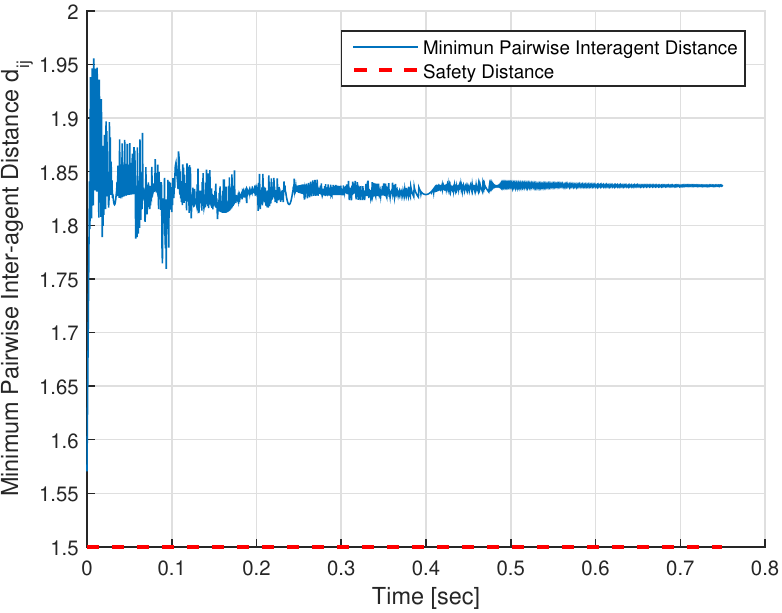}
\caption{The minimum inter-agent distance between any two robots for a simulated trial with $N=15$. Note that as this is always greater than the allowable safety distance, there were never any collisions between robots.}
\label{fig:noncollidingPaths15}
\end{figure}

\begin{figure}
\centering
\includegraphics[width=0.975\columnwidth,clip]{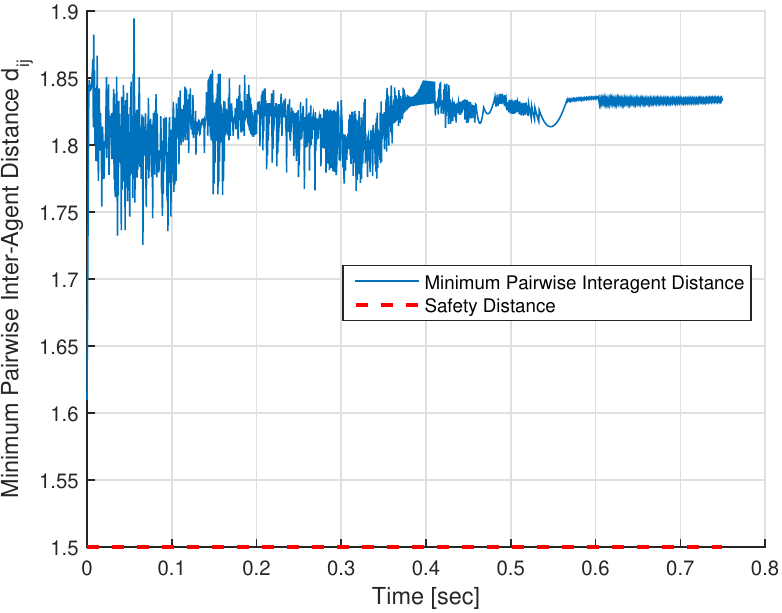}
\caption{The minimum inter-agent distance between any two robots for a simulated trial with $N=40$. Note that as this is always greater than the allowable safety distance, there were never any collisions between robots.}
\label{fig:noncollidingPaths40}
\end{figure}

The initial positions and the assigned goal locations are selected in very close proximity so that the \ac{OGA} policy on its own is not sufficient to ensure collision-free trajectories, see Fig. \ref{fig:15RobotSim} and \ref{fig:40RobotSim}, respectively. The agents start from the initial conditions shown in Fig. \ref{fig:init15}, \ref{fig:init40} towards goal locations marked with ``x''; as discussed earlier, the initial goal assignment is not necessarily the optimal one. The communication ranges are denoted by the rings around each agent, and set equal to $R_c = 4 R$, where $R$ is the radius of the agents. The straight line paths from the starting locations to the initially assigned goals are depicted as the dotted lines, whereas the actual paths followed by the switching strategy are shown in Fig. \ref{fig:paths15} and \ref{fig:paths40}, respectively. The final locations of the robots are shown in Fig. \ref{fig:final15} and \ref{fig:final40}, respectively. As indicated by the colors of the initially assigned goals, the followed paths, and the final goal locations of the robots, the robots performed goal swaps during the duration of the simulation under the \ac{OGA} policy, while they also moved under the \ac{CA} policy during the path segments that are not straight. Figures \ref{fig:noncollidingPaths15} and \ref{fig:noncollidingPaths40}, respectively, show the minimum distance between any two pairs of agents at each time instance of the simulations. As this distance is always greater than the safety distance $d_s$, there is never a collision between any robots.

Additional videos with teams of 6, 30, 80, 100, and 150 agents are available through the following links:

\noindent https://www.dropbox.com/s/a5hykza2wla7x3n/6.mp4?dl=0
https://www.dropbox.com/s/1gazjlt7nell0mn/30.avi?dl=0
https://www.dropbox.com/s/48bnwsnlqibkgaj/80.avi?dl=0
https://www.dropbox.com/s/e6wljow9ft2ziap/100.avi?dl=0
https://www.dropbox.com/s/uif0yahjlhv0bsf/150.avi?dl=0

It is noteworthy that the simulation videos demonstrate the expected performance of the overall algorithm in terms of the convergence rate; more specifically, the slower convergence of the agents to their goal locations towards the end of the simulation, compared to their faster motion in the beginning of the simulation, is consistent with the exponential convergence rate of the \ac{OGA} policy. \footnote{Note however that the videos run at different frames-per-second or have deliberate pauses to explain the algorithm, and as thus the time lengths of the videos shall not be used for a direct comparison of the convergence time among the cases of different $\N$.} 

\subsection{Comparison with earlier work in \cite{Panagou_Turpin_Kumar_ICRA14}}

\begin{figure}
{
\centering
\subfigure[Initial positions (red), goal locations (green) and the paths resulting from the current approach.]{
\includegraphics[width=0.8\columnwidth,clip]{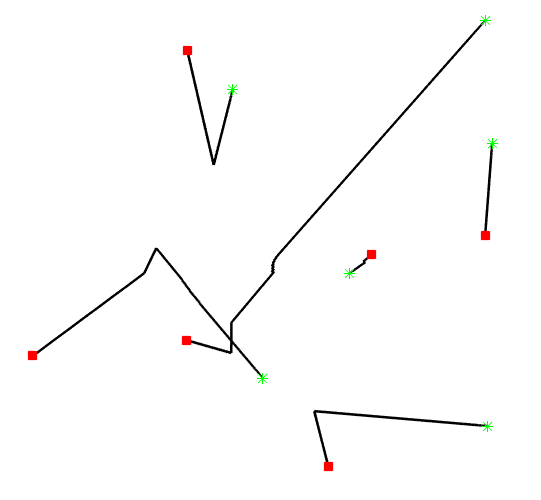}\label{fig:new}
}
\subfigure[Initial positions (red), goal locations (green) and the paths resulting from the approach in \cite{Panagou_Turpin_Kumar_ICRA14}.]{
\includegraphics[width=0.8\columnwidth,clip]{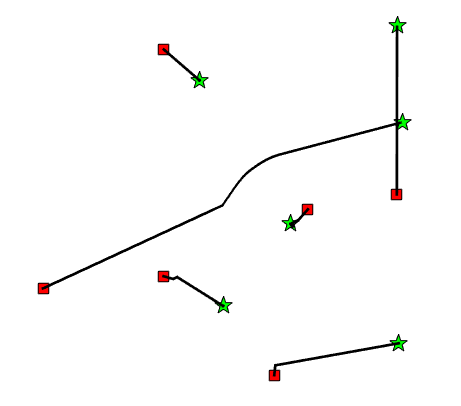}\label{fig:icra}
}
}
\caption{Comparison of the paths resulting from the current approach versus the approach in \cite{Panagou_Turpin_Kumar_ICRA14}.}
\label{fig:comparison}
\end{figure}

As discussed earlier, the current approach dominates our earlier work \cite{Panagou_Turpin_Kumar_ICRA14} in the following aspects: (i) The initial positions of the robots and the goal locations can be close as $d_s= 2R$, in contrast to \cite{Panagou_Turpin_Kumar_ICRA14} that requires the initial locations and goals to be $2R\sqrt{2}$ apart for safety. (ii) The collision-free trajectories of the current approach are rendered via switching among two control policies, opposed to \cite{Panagou_Turpin_Kumar_ICRA14} that involves switching among three control policies. The approach here nevertheless does not consider the initially optimal assignment, and as thus, depending on the connected subcomponents at initial time and throughout the evolution of the system, might result in suboptimal paths in terms of total distance. For completeness, we include simulation results between the current approach and \cite{Panagou_Turpin_Kumar_ICRA14}; in the scenario illustrated in Fig. \ref{fig:comparison}, the resulting paths are suboptimal; the total computational time for the entire simulation in the former case is was 2.625 sec, while in the latter case was 1.375 sec. This demonstrates that even with suboptimal paths, the algorithm is computationally efficient for the considered applications.           

\subsection{Notes on communication requirements and planning time}\label{Comm-req}

The implementation of the proposed algorithm in a realistic setting requires the capability of exchanging messages (e.g., via wireless communication links) among the subset of agents that are considered connected at some time instance, i.e., among robots that lie within $R_c$ distance and their neighbor agents. Once a connected group is formed, the \ac{OGA} policy runs in a synchronous manner. Recall that in our approach the initial assignment is \emph{random}, i.e., the focus of this work is not on determining the globally optimal assignment of the robots to the goals, but instead we focus on re-assigning goals among connected robots. Thus, since the decision-making mechanism (i.e., the \ac{OGA} policy) on re-determining the goal assignment is triggered whenever (a subset of) agents become connected, it follows that the number of re-plan operations, and hence the number of messages that are exchanged among the connected agents on deciding whether they swap goals, depends greatly on the communication range, the number of robots, and their initial distribution, i.e., the quality (in terms of sub-optimality) of the initial assignment. Note that for sufficiently large radius $R_c$, which results in all agents being connected at the same time, the centralized case is recovered, meaning that the agents decide among $\N!$ assignments. 

The time for a re-plan operation also directly depends on the number of agents in the connected component. Planning times on simulated teams of $\N_c$ robots using a single computer take approximately $10^{-7}{\N_c}^3$ seconds, which reads that a component of 100 connected robots takes about 0.1 seconds to re-plan. The required computational time is typically sufficient for the proposed multi-robot applications.

\section{Experimental Results}\label{Experimental Results}

The proposed switching logic and control policies were also evaluated in experimental trials with six ground robots (AION R1 Rovers) in the Distributed Aerospace Systems and Control Lab, University of Michigan. The position of the robots was acquired via a VICON motion capture system. The Robot Operating System (ROS) was used to exchange data among the agents, as well as to convert the high-level control commands to low-level motor voltage commands that drive the robots' wheels. The control loop was closed at approximately every 30 msec. Since the robots are differentially-driven, hence closer modeled via the unicycle nonholonomic kinematics rather than single integrators, the proposed controllers were implemented by steering the robots' orientations along the direction of the velocity vector computed from either the control law \eqref{controller} or the control law \eqref{uj}, and applying the corresponding linear velocity along this direction. This corresponds in part to the discrepancy in the resulting paths between the simulation results in Fig. \ref{fig:new} and experimental results as described below.  

The experiments demonstrate that the proposed switching logic and corresponding policies are applicable in practice, and produce results that are consistent with the expected theoretical performance. The controller parameters were set equal to: $R=0.35$ m (robot radius), $R_c = 1.1$ m (communication radius), $R_s=0.9$ m, $d_s=0.7$ m (minimum allowable distance). The video of experiment is available at 

\noindent https://www.dropbox.com/s/ujrtzn1aolc0yul/6-exp.mp4?dl=0.

\begin{figure}
{
\centering
\subfigure[The resulting paths to the goals under the proposed control strategy.]{
\includegraphics[width=0.95\columnwidth,clip]{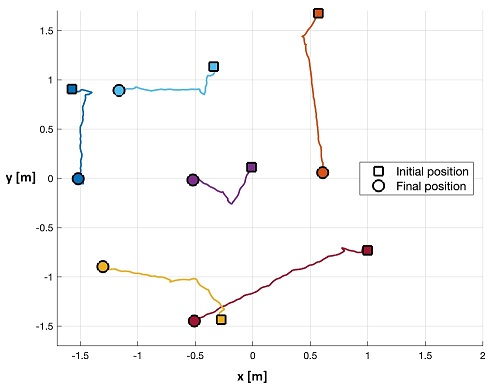}\label{fig:paths-exp}
}
\subfigure[The distances of the agents w.r.t. their goals over time, and the invoked goal swaps when the \ac{OGA} becomes active.]{
\includegraphics[width=0.95\columnwidth,clip]{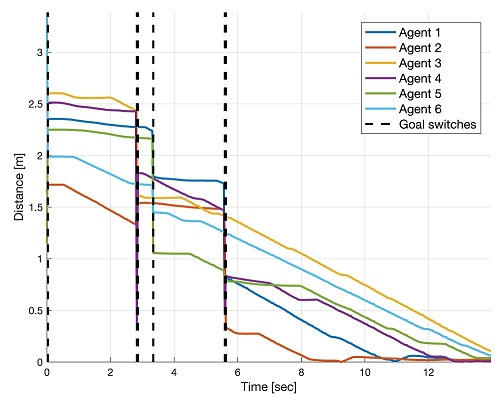}\label{fig:goals-exp}
}
\subfigure[The minimum clearance between any two robots over time.]{
\includegraphics[width=0.95\columnwidth,clip]{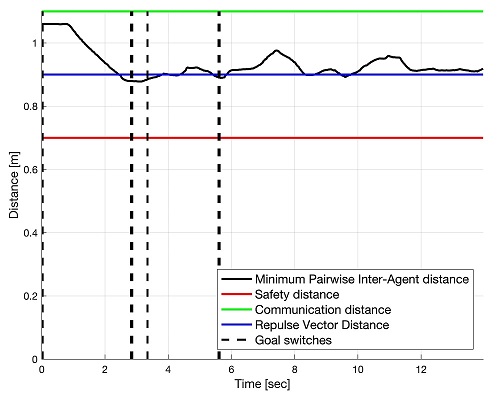}\label{fig:dist-exp}
}
}
\caption{Experimental results with $N=6$ robots.}
\label{fig:experiments}
\end{figure}

The derived paths of the robots under the proposed approach are shown in Fig. \ref{fig:paths-exp}. Fig. \ref{fig:goals-exp} depicts the distance of each robot \ac{wrt} its corresponding goal over time, and furthermore illustrates the time instances where the \ac{OGA} becomes active and invokes goal swaps along with concurrent reduction in the cost (distance-to-the-goal). The minimum pairwise distance among the robots over time is shown in Fig. \ref{fig:dist-exp}, and demonstrates that the trajectories are always collision-free.

\section{Concluding Remarks}\label{Conclusions}
We presented a switched systems approach on the decentralized concurrent goal assignment and trajectory generation for multi-robot networks which guarantees safety and global stability to interchangeable goal locations. The proposed switching logic relies on multiple Lyapunov-like functions that encode goal swap among locally connected agents based on the total traveled distance, avoidance of inter-agent collisions, and convergence to the assigned goal locations. As such, the proposed methodology renders feedback control policies with local coordination only, and therefore is suitable for applications such as in first-response deployment of robotic networks under limited information sharing. The proposed algorithms were evaluated via simulation results and experiments with six robots. Our future work will focus on agents with more complicated dynamics, as well as on the robustness of our algorithms under communication failures and uncertainty.

\bibliographystyle{IEEEtran}
\bibliography{references}

\begin{biography}[{\includegraphics[width=1in,height=1.25in,clip,keepaspectratio]{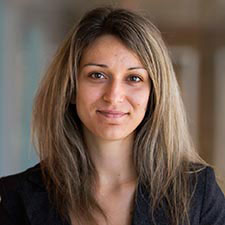}}]{Dimitra Panagou} received the Diploma and PhD degrees in Mechanical Engineering from the National Technical University of Athens, Greece, in 2006 and 2012, respectively. Since September 2014 she has been an Assistant Professor with the Department of Aerospace Engineering, University of Michigan. Prior to joining the University of Michigan, Dr. Panagou was a postdoctoral research associate with the Coordinated Science Laboratory, University of Illinois, Urbana-Champaign (2012-2014), a visiting research scholar with the GRASP Lab, University of Pennsylvania (June 2013, fall 2010) and a visiting research scholar with the University of Delaware, Mechanical Engineering Department (spring 2009). Her research interests include the fields of planning, coordination, and distributed control and estimation for complex systems, with applications in unmanned aerial systems, robotic networks and autonomous multi-vehicle systems (ground, marine, aerial, space). She was a recipient of the NASA 2016 Early Career Faculty Award and of the Air Force Office of Science Research 2017 Young Investigator Award.
\end{biography}

\begin{biography}[{\includegraphics[width=1in,height=1.25in,clip,keepaspectratio]{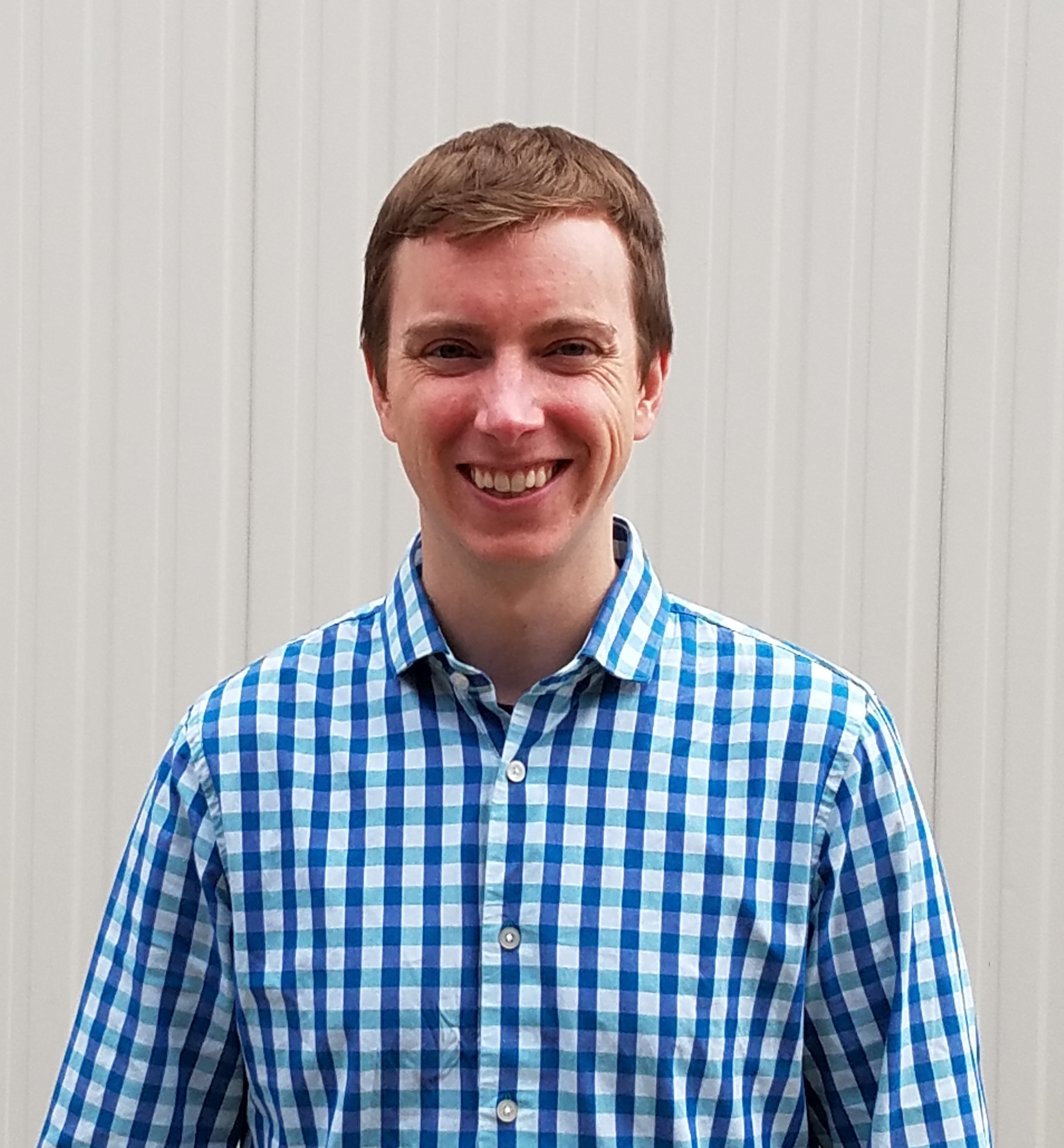}}]{Matthew Turpin} received his BS degree in Mechanical Engineering from Northwestern University in 2009, and his PhD degree in Mechanical Engineering from University of Pennsylvania in 2014. During his doctoral studies at the University of Pennsylvania he focused on formation control as well as on the coupling of task assignment and motion planning in multi-robot systems. He is currently a research scientist at Qualcomm Technologies Inc. where he has worked on research and development of software for multi-rotor MAVs, and currently works on planning and control for autonomous cars.
\end{biography}

\begin{biography}[{\includegraphics[width=1in,height=1.25in,clip,keepaspectratio]{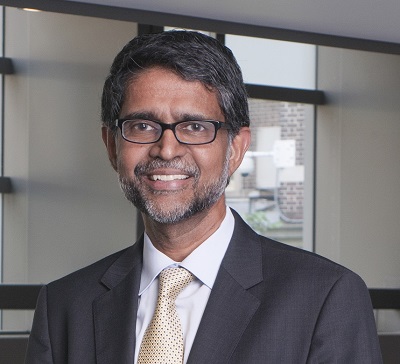}}] {Vijay Kumar} is the Nemirovsky Family Dean of Penn Engineering with appointments in the Departments of Mechanical Engineering and Applied Mechanics, Computer and Information Science, and Electrical and Systems Engineering at the University of Pennsylvania. He served as the assistant director of robotics and cyber physical systems at the White House Office of Science and Technology Policy (2012-2014) and is the founder of Exyn Technologies, a company that develops solutions for autonomous flight. 

Dr. Kumar is a Fellow of the American Society of Mechanical Engineers (2003), a Fellow of the Institute of Electrical and Electronic Engineers (2005) and a member of the National Academy of Engineering (2013). Among his recent awards are the 2012 World Technology Network award, a 2013 Popular Science Breakthrough Award, a 2014 Engelberger Robotics Award, the 2017 IEEE Robotics and Automation Society George Saridis Leadership Award in Robotics and Automation and the 2017 ASME Robert E. Abbott Award.  He has won best paper awards at DARS 2002, ICRA 2004, ICRA 2011, RSS 2011, RSS 2013, and BICT 2015 and has advised doctoral students who have won Best Student Paper Awards at ICRA 2008, RSS 2009, and DARS 2010.\end{biography}

\end{document}